\documentclass[aps,twocolumn,showpacs,showkeys,superscriptaddress]{revtex4}
\usepackage{epsfig}
\usepackage{mathrsfs}
\usepackage{graphicx}

\newcommand{\bastar}{\begin{eqnarray*}}
\newcommand{\eastar}{\end{eqnarray*}}
\newskip\humongous \humongous=0pt plus 1000pt minus 1000pt

\newif\ifdtup

\relax
\newcommand{\bea}{\begin{eqnarray}}
\newcommand{\eea}{\end{eqnarray}}
\newcommand{\X}{{\vec X}}
\newcommand{\pro}{\partial}
\newcommand{\n}{\hat n}
\newcommand{\mn}{{\mu\nu}}
\newcommand{\oneg}{\displaystyle\frac{1}{g}}

\newcommand{\bq}{\bar q}

\newcommand{\bs}{\bar s}
\newcommand{\br}{\bar r}
\newcommand{\bb}{\bar b}
\newcommand{\bg}{\bar g}
\newcommand{\bR}{\bar R}
\newcommand{\bB}{\bar B}
\newcommand{\bG}{\bar G}
\newcommand{\F}{\vec F}
\newcommand{\W}{\vec W}

\newcommand{\hF}{\hat F}

\newcommand{\A}{{\vec A}}
\newcommand{\hA}{{\hat A}}
\newcommand{\cA}{{\cal A}}
\newcommand{\cC}{{\cal C}}
\newcommand{\valpha}{{\vec \alpha}}

\newcommand{\hn}{{\hat n}}
\newcommand{\hD}{{\hat D}}
\newcommand{\dfrac}{\displaystyle\frac}

\newcommand{\nn}{\nonumber}

\begin{document}
\title{Glueball Physics in QCD}
\bigskip
\author{Y. M. Cho}
\email{ymcho7@konkuk.ac.kr}
\affiliation{Institute of Modern Physics, Chinese Academy of
Science, Lanzhou 730000, China}
\affiliation{Administration Building 310-4,
Konkuk University, Seoul 143-701, Korea}
\affiliation{School of Physics and Astronomy,
Seoul National University, Seoul 151-747, Korea}
\author{X. Y. Pham}
\affiliation{Lab de Physique Theorique et Hautes Energies \\
Tour 24-25, 1er Etage, 4 Place Jussieu, Universite Paris VI, 
F-75252 Paris, France}
\author{Pengming Zhang}
\affiliation{Institute of Modern Physics, Chinese Academy of
Science, Lanzhou 730000, China}
\affiliation{State Key Laboratory of
Theoretical Physics, Institute of Theoretical Physics, Chinese
Academy of Sciences, Beijing 100190, China}
\author{Ju-Jun Xie}
\affiliation{Institute of Modern Physics, Chinese Academy of
Science, Lanzhou 730000, China}
\affiliation{State Key Laboratory of
Theoretical Physics, Institute of Theoretical Physics, Chinese
Academy of Sciences, Beijing 100190, China}
\author{Li-Ping Zou}
\affiliation{Institute of Modern Physics, Chinese Academy of
Science, Lanzhou 730000, China}
\affiliation{State Key Laboratory of
Theoretical Physics, Institute of Theoretical Physics, Chinese
Academy of Sciences, Beijing 100190, China}
\affiliation{Institute of Basic Sciences, Konkuk University,
Seoul 151-747, Korea}

\date{\today}

\begin{abstract}

The Abelian decomposition of QCD which decomposes the 
gluons to the color neutral binding gluons and the colored 
valence gluons shows that QCD can be viewed as the restricted 
QCD (RCD) made of the binding gluons which has the valence 
gluons as colored source, and simplifies the QCD dynamics 
greatly. In particular, it tells that the gauge covariant valence 
gluons can be treated as the constituents of hadrons, and 
generalizes the quark model to the quark and valence gluon 
model. So it provides a comprehensive picture of glueballs 
and their mixing with quarkoniums, and predicts new hybrid 
hadrons made of quarks and valence gluons. We discuss 
how these predictions could be confirmed experimentally. 
In particular we present a systematic search for the ground 
state glueballs and their mixing with quarkoniums below 
2 GeV in $0^{++}$, $2^{++}$, and $0^{-+}$ channels within 
the framework of QCD, and predict the relative branching 
ratios of the radiative decay of $\psi$ to the physical states. 

\end{abstract}

\maketitle

\section{Introduction}

One of the important issues in hadron spectroscopy is the 
identification of glueballs. The general wisdom is that QCD 
must have the glueballs made of gluons. In early days the 
gauge invariant combinations of the QCD field strength were 
suggested to generate the glueballs \cite{frit,pf,kogut}. 
Later several models of glueballs including the bag model 
and the constituent model have been proposed \cite{jaff,roy,coyne,chan,corn,ish,ams,math,ochs}. 
Moreover, the lattice QCD has been able to estimate the mass 
of the low-lying glueballs \cite{lqcd1,lqcd2,lqcd3,lqcd4}. 

But so far the search for the glueballs has not been so successful
for two reasons. First, theoretically there has been no consensus 
on how to construct the glueballs from QCD. For example, there 
has been the proposal to make the glueballs with ``the constituent 
gluons", but a precise definition of the constituent gluon was 
lacking \cite{coyne,chan}. This has made it difficult for us 
to predict what kind of glueballs we can expect.

The other reason is that it is not clear how to identify
the glueballs experimentally. This is partly because the
glueballs could mix with the quarkoniums, so that we must
take care of the possible mixing to identify the glueballs
experimentally \cite{coyne,chan}. This is why we have 
very few candidates of the glueballs so far, compared to 
huge hadron spectrum made of quarks listed by Particle Data 
Group (PDG) \cite{pdg}.

This makes the search for the glueballs an important issue 
in hadron spectroscopy. Indeed, one of the main purpose of 
the Jefferson Lab 12 GeV upgrading is to search for the 
glueballs \cite{jlab}. The purpose of this paper is to provide 
a comprehensive and clear picture of the glueballs in QCD, 
to study the possible mixing with the quarkoniums, and to 
discuss how to identify them theoretically and experimentally.

Actually, it is not difficult to define the gauge covariant 
colored gluons which form color octet which could be identified 
as the constituent gluons. This can be done with the Abelian 
decomposition known as the Cho decomposition or Cho-Duan-Ge 
(CDG) decomposition, which decomposes the QCD gauge potential 
to the color neutral restricted potential and the colored valence 
potential gauge independently \cite{prd80,prl81,prd81,prd02a,duan}.

What is remarkable about this decomposition is that the restricted
potential has the full non-Abelian gauge symmetry and the valence
potential transforms gauge covariantly. So we can construct the
restricted QCD (RCD) which describes the core dynamics of QCD 
with the restricted potential, and view QCD as RCD which has the 
valence gluons as the gauge covariant colored source. 

Clearly the Abelian decomposition tells that there are two types 
of gluons which play different roles. The restricted potential 
describes the color neutral binding gluons which confines the 
colored source, and the valence potential describes the colored 
valence gluons which become the colored source of QCD. 

This seems to justify the intuitive idea of the constituent 
gluons, because the valence gluons can be viewed as the 
constituents of hadrons. But there is an important difference.
Here the binding gluons are not treated as the constituents. 
To understand this consider the atomic bound states in QED. 
Obviously we have photons as well as electrons (and protons) 
in atoms, but only the electron and proton become the 
constituents because only they determine the atomic structure
of the periodic table. The photons play no role in the periodic 
table. They are there as the electromagnetic field to provide 
the binding force and binding energy, not as a constituent 
particle to determine the atomic structure.  

Exactly the same way we need quarks and gluons to make the
proton. But again the gluons inside the proton does not play 
any role in the baryon spectrum. This means that they must 
be the ``binding" gluons, not the ``constituent" gluons, which 
(just like the photons in atoms) provide only the binding force 
and the binding energy of the proton. If so, what are the binding 
gluons and the constituent gluons? And how can one distinguish 
them? This is the problem of the constituent model. 

Clearly the Abelian decomposition provides a natural answer. 
It tells that there are indeed two types of gluons, binding 
gluons and valence gluons, and only the valence gluons 
can be treated as the constituent gluons. And the gluons in 
proton are the binding gluons, not the valence gluons, because 
they play no role to determine the position of the proton in 
the baryon spectrum. Only three constituent quarks characterize 
the baryonic structure of the proton.

This tells that the binding gluons can not be the constituent
of hadrons. As importantly this tells that we can treat the 
colored valence gluons, just like the quarks, as the constituent 
particles in QCD. In particular, we can easily consrtuct the 
color singlet glueballs with two or three valence gluons. This 
provides a clear picture of the glueballs in QCD and helps us 
to identify the glueball more clearly \cite{prd80,prl81,prd81}.

A potential problem with this picture of the glueballs is that 
this could give us too many glueballs, while experimentally 
we have few candidates of them so far. This is a big mystery 
in hadron spectroscopy. So the real problem with the glueballs 
is to understand why there are so few candidates of them 
experimentally, compared to the rich hadron spectrum based 
on the successful quark model. 

As we have already mentioned, one reason (at least partly) 
is the possible mixing with the quarkoniums. This makes the 
experimental identification of glueballs a non-trivial matter.
To resove this problem we need a clear picture of the mixing 
mechanism, and the Abelian decomposition can easily provide 
this \cite{snu89}. 

Another reason is that the glueballs have an intrinsic 
instability. To understand this we must understand the 
confinement mechanism in QCD more clearly, and the Abelian 
decomposition provides this \cite{prd13,ijmpa14}. First, 
it assures that only the restricted potential can contribute 
to the Wilson loop integral \cite{prd00}. This can easily 
be understood because the valence gluons (being colored) 
become the confined prisoners, so that only the binding 
gluons can be the confining agents \cite{prl81,prd81}. 
This, of course, is the Abelian dominance \cite{thooft}. 

However, the Abelian dominance does not tell what is the 
confinement mechanism. This is because the restricted 
potential is made of two parts,  the non-topological Maxwell 
part and the topological Dirac's monopole part \cite{prd80,prl81}. 
And the Abelian dominance does not tell which part generates 
the confinement, and how.
 
Fortunately we can tell which part is responsible for the 
confinement. Implementing the Abelian decomposition on 
lattice we can calculate the Wilson loop numerically with 
the full potential, the restricted potential, and the monopole 
potential separately, and show that the monopole potential 
is responsible for the area law in the Wilson loop gauge 
independently \cite{kondo1,kondo2,cundy1,cundy2}. 

Moreover, we can tell that it is the monopole condensation, 
more precisely the monopole-antimonopole pair condensation,
which generates the confinement in QCD. The Abelian 
decomposition allows us to calculate the QCD effective 
action in the presence of the monopole background and 
establish the stable monopole condensation gauge 
independently \cite{prd13,ijmpa14}. This tells that the 
true vacuum of QCD is given by the monopole condensation 
which generates the dimensional transmutation and the mass 
gap.

This picture of the confinement helps us to understand 
why there are not so many candidates of the glueballs,
because this tells that the glueballs made of the valence 
gluons have an intrinsic instability. The effective action of 
QCD tells that the colored gluons, unlike the quarks, tend 
to annihilate each other in the chromo-electric background. 
This must be contrasted with quarks, which remain stable 
inside the hadrons. The reason is that the chromo-electric 
field tend to create the quark pairs, but annihilate the valence 
gluons \cite{sch,prd02b,jhep05}.

This is closely related to the asymptotic freedom (anti-screening)
of gluons. It is well known that in QED the strong electric background 
tends to generate the pair creation of electrons, which makes the 
charge screening \cite{schw,prl01}. But in QCD gluons and quarks 
play opposite roles in the asymptotic freedom. The quarks enhance 
the screening while the gluons (overide the quarks and) diminish 
it to generate the anti-screening \cite{wil,pol}. We can understand 
this with the pair creation of the quarks and the pair annihilation 
of the valence gluons in the chromo-electric field \cite{prd13,ijmpa14}.

Clearly the Abelian decomposition predicts new hybrid hadrons 
made of quarks and valence gluons, in addition to the above 
glueballs. This is because the valence gluons, just like the 
quarks, can be viewed as the constituents of (not just the 
glueballs but) the hadrons. This suggests us to generalize 
the quark model to the ``quark and valence gluon" model, in 
which both quarks and valence gluons become the constituents 
of hadrons \cite{prd80,prl81,prd81}. 

In this generalization of the quark model we can construct color 
singlet hadrons from the valence gluons and the valence quarks. 
For example, we can have a $q\bq g$ color singlet hybrid meson 
with one octet valence gluon $g$ and a $q \bq$ octet. Or, we 
can have a $qqq g$ hybrid baryon from the $qqq$ octet and 
one valence gluon octet $g$. So the Abelian decomposition 
of QCD providea a totally new picture of hadron spectroscopy.  

Of course, there have been proposals of hybrid hadrons 
before \cite{barnes,isg,hyb1,hyb2}. But a clear picture 
of hybrid hadrons was missing. The quark and valence gluon 
model provides a clear picture, and helps us to identify 
them experimentally. 

Finally, the above picture of confinement predicts a totally 
different type of glueball, the ``magnetic" glueball which 
we can call the ``monoball" \cite{prl81,prd81}. This is because 
the monopole condensation could most likely have the vacuum 
fluctuation which can naturally be identified as a $0^{++}$ 
state, which represents the mass gap generated by the monopole 
condensation. This is the monoball. Clearly this has nothing 
to do with the above glueballs made of the valence gluons.

The importance of the monoball comes from the fact that it 
is a direct consequence of the monopole condensation. So 
the identification of the monoball could be interpreted as the 
experimental confirmation of the monopole condensation 
in QCD. This makes the experimental verification of the 
monoball a most urgent issue in QCD. 

Although the quark model has been very successful, PDG tells 
that there are experimentally established hadronic states which 
can not easily be explained by the quark model. For example, 
the scalar meson $f_0(500)$ or $f_0(980)$ does not seem to fit 
in the simple quark model, although there have been many efforts 
to explain this within the quark model \cite{mor,jaffe2,wein}. 
We hope that our analysis in this paper will help to identify 
their physical content more clearly. 

The paper in organized as follows. In Section II we review 
the Abelian decomposition and the confinement mechanism 
for later purpose. In Section III discuss the glueball spectrum 
in QCD. In Section IV we discuss the glueball-quarkonium 
mixing. In Section V we present the numerical analysis of the 
low-lying glueball-quarkonium mixing in $0^{++}$, $2^{++}$, 
and $0^{-+}$ channels. In Section VI we briefly discuss the 
hybrid hadrons in QCD. In Section VII we discuss the monoball
as the experimental evidence of the monopole condensation 
in QCD. Finally in the last section we discuss the physical 
implications of our analysis.    

\section{Binding Gluons and Valence Gluons: A Review} 

It is well known that QCD can be understood as the extended 
QCD (ECD), namely RCD made of the binding gluons which has 
the valence gluons as colored source \cite{prd80,prl81,prd81}. 
This follows from the Abelian decomposition of QCD which 
decomposes the gauge potential to the restricted part and 
the valence part gauge independently.

To show this we review the Abelian decomposition first. Conside 
the SU(2) QCD for simplicity, and let $(\hn_1,\hn_2,\hn_3=\hn)$ 
be an arbitrary right-handed local orthonormal basis. To make 
the Abelian decomposition we choose $\hn$ to be the Abelian 
direction, and impose the isometry to project out the restricted 
potential $\hA_\mu$ which describes the Abelian sub-dynamics of 
QCD \cite{prd80,prl81,prd81}
\bea
&D_\mu \hn=(\pro_\mu+g\A_\mu \times) \hn=0,  \nn\\
&\A_\mu \rightarrow \hA_\mu=A_\mu \n-\oneg \n \times \pro_\mu \n
=\cA_\mu+\cC_\mu,  \nn\\
&\cA_\mu=A_\mu \n,~~\cC_\mu=-\oneg \n \times \pro_\mu \n,
~~A_\mu=\hn \cdot \A_\mu. 
\label{ap}
\eea
This is the Abelian projection which projects out the color neutral
binding gluons. Notice that $\hA_\mu$ is precisely the connection
which leaves the Abelian direction invariant under the parallel
transport. Remarkably, it is made of two parts, the topological
(Diracian) $\cC_\mu$ which describes the non-Abelian monopole 
as well as the non-topological (Maxwellian) $\cA_\mu$.

Moreover, we have
\bea
& \hat{F}_{\mu\nu} = (F_{\mu\nu}+ H_{\mu\nu})\hn
=G_\mn \n, \nn\\
&F_\mn = \pro_\mu A_\nu-\pro_\nu A_\mu,  \nn\\
&H_\mn = \partial_\mu C_\nu-\partial_\nu C_\mu,
~~~C_\mu=-\dfrac{1}{g} \hn_1 \cdot \pro_\mu \hn_2, \nn\\
&G_\mn = \pro_\mu B_\nu-\pro_\nu B_\mu, 
~~~B_\mu=A_\mu+C_\mu.
\label{rf} 
\eea 
This tells two things. First, $\hF_\mn$ has only the Abelian 
component. Second, $\hF_\mn$ is made of two potentials, the
electric (non-topological) $A_\mu$ and magnetic (topological)
$C_\mu$. 

Under the infinitesimal gauge transformation 
\bea
\delta \A_\mu=\oneg  D_\mu \vec \alpha, 
~~~\delta \hn_i= -\vec \alpha \times \hn_i, 
\eea 
we have 
\bea 
&\delta A_\mu= \dfrac1g \hn\cdot \pro_\mu \vec \alpha, 
~~~\delta C_\mu= -\dfrac1g \hn\cdot \pro_\mu \vec \alpha, 
\eea 
so that
\bea 
&\delta \hA_\mu=\dfrac1g \hD_\mu \vec \alpha, 
~~~(\hD_\mu=\pro_\mu+g\hA_\mu\times).
\eea
This tells that $\hA_\mu$ has the full SU(2) gauge degrees of
freedom, even though it is restricted.

From this we can construct RCD which has the full non-Abelian
gauge symmetry but is simpler than the QCD 
\bea 
&{\cal L}_{RCD} =-\dfrac{1}{4} \hF^2_\mn
=-\dfrac{1}{4} F_\mn^2 \nn\\
&+\dfrac1{2g} F_\mn \hn \cdot (\pro_\mu \hn \times \pro_\nu \hn)
-\dfrac1{4g^2} (\pro_\mu \hn \times \pro_\nu \hn)^2,
\label{rcd}
\eea
which describes the Abelian subdynamics of QCD. Since RCD
contains the non-Abelian monopole degrees explicitly, it
provides an ideal platform for us to study the monopole
dynamics gauge independently.

With (\ref{ap}) we can recover the full QCD potential adding
the non-Abelian (colored) part $\X_\mu$ \cite{prd80,prl81,prd81}
\bea
&\A_\mu = \hA_\mu + \X_\mu,     \nn\\
&\X_\mu=\dfrac1g \hn \times D_\mu \hn,
~~~~\hn \cdot \X_\mu=0.
\label{adec}
\eea
Under the gauge transformation we have
\bea
&\delta \hA_\mu = \oneg  \hD_\mu \vec \alpha,
~~~\delta \X_\mu = - \valpha \times \X_\mu.
\label{cgt}
\eea
This confirms that $\vec X_\mu$ becomes gauge covariant.
This is the Abelian decomposition which decomposes the gluons 
to the color neutral binding gluons and the colored valence 
gluons gauge independently. This is known as Cho decomposition, 
Cho-Duan-Ge (CDG) decomposition, or Cho-Faddeev-Niemi (CFN) 
decomposition \cite{fadd,shab,gies,zucc}.

From (\ref{adec}) we have
\bea
\vec{F}_{\mu\nu}&=&\hat F_{\mu \nu} + \hD _\mu \X_\nu - \hD_\nu
\X_\mu + g\X_\mu \times \X_\nu. 
\eea 
With this we can express QCD by
\bea 
&{\cal L}_{QCD} = -\dfrac{1}{4} \F^2_{\mu\nu }
=-\dfrac{1}{4}\hF_\mn^2-\dfrac{1}{4}(\hD_\mu\X_\nu
-\hD_\nu\X_\mu)^2 \nn\\
&-\dfrac{g}{2} {\hat F}_{\mu\nu} \cdot (\X_\mu \times \X_\nu)
-\dfrac{g^2}{4} (\X_\mu \times \X_\nu)^2. 
\label{ecd} 
\eea
This is the extended QCD (ECD) which confirms that QCD can be 
viewed as RCD made of the binding gluons, which has the colored 
valence gluons as its source \cite{prd80,prl81,prd81,prd02a}. 

\begin{figure}
\begin{center}
\includegraphics[scale=0.6]{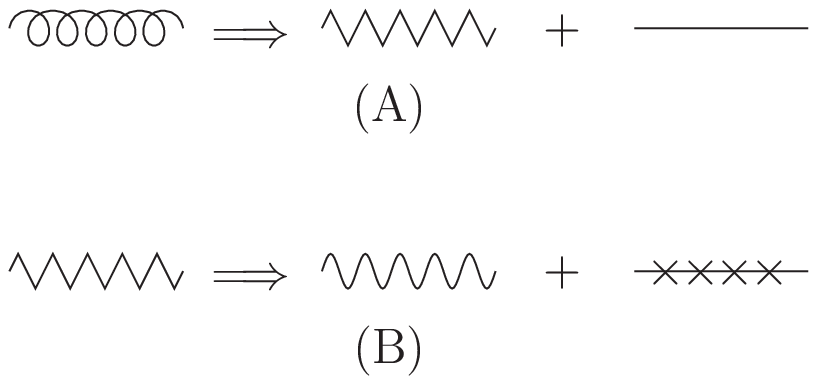}
\caption{\label{cdec} The Abelian decomposition of the gluons.
The gluon is decomposed to the binding gluon (kinked line) and
the valence gluon (straight line) in (A), and the binding gluon
is further decomposed to the Maxwell part (wiggly line) and
Dirac part (spiked line) in (B).}
\end{center}
\end{figure}

We can express the Abelian decomposition of the gluons given by 
(\ref{ap}) and (\ref{adec}) graphically. This is shown in Fig. \ref{cdec},
where the gluons are decomposed to the binding gluons and the
valence gluons in (A), and the binding gluons are decomposed
further to the non-topological Maxwell part $\cA_\mu$ and
the topological Dirac part $\cC_\mu$ in (B). 

The Abelian decomposition of SU(3) QCD is a bit more
complicated, but is well known. Since SU(3) has rank two, 
we have two Abelian subgroups in SU(3). Let $\n_i~(i=1,2,...,8)$ 
be the local orthonormal SU(3) basis. Clearly we can choose the 
Abelian directions to be $\n_3=\n$ and $\n_8=\n'$. Now make 
the Abelian projection by
\bea
D_\mu \n=0.
\label {3cp}
\eea
This automatically guarantees \cite{prl80}
\bea
D_\mu \n'=0,~~~\n'=\dfrac1{\sqrt 3} \n*\n.
\eea
where $*$ denotes the $d$-product. This is because SU(3) has two
vector products, the anti-symmetric $f$-product and the symmetric
$d$-product.

Solving (\ref{3cp}), we have the following Abelian projection
which projects out two neutral binding gluons,
\bea
&\A_\mu \rightarrow \hA_\mu=A_\mu \hn+A_\mu' \hn'
-\oneg \hn\times \pro_\mu \hn-\oneg \hn'\times \pro_\mu \hn' \nn\\
&=\sum_p \dfrac23 \hA_\mu^p,~~~(p=1,2,3),    \nn\\
&\hA_\mu^p=A_\mu^p \n^p-\oneg \n^p \times \pro_\mu \n^p
=\cA_\mu^p+\cC_\mu^p, \nn\\
&A_\mu^1=A_\mu,
~~~A_\mu^2=-\dfrac{1}{2}A_\mu+\dfrac{\sqrt 3}{2}A_\mu',  \nn\\
&A_\mu^3=-\dfrac{1}{2}A_\mu-\dfrac{\sqrt 3}{2}A_\mu',
~~~\n^1=\n,  \nn\\
&\n^2=-\dfrac{1}{2} \n +\dfrac{\sqrt 3}{2} \n',
~~~\n^3=-\dfrac{1}{2} \n -\dfrac{\sqrt 3}{2} \n',
\label{cp3}
\eea
where the sum is the sum of the Abelian directions of three SU(2)
subgroups made of $(\n_1,\n_2,\n^1),~(\n_6,\n_7,\n^2),~(\n_4,-\n_5,\n^3)$.
Notice the factor $2/3$ in front of $\hA_\mu^p$ in the $p$-summation.
This is because the three SU(2) binding potentials are not independent.

From this we have the restricted field strength
\bea
&\hF_\mn=\sum_p \dfrac23 \hF_\mn^p,
\eea
which is made of two Abelian binding potentials. With this we have
the restricted QCD
\bea
&{\cal L}_{RCD} = -\sum_p \dfrac{1}{6} (\hF_\mn^p)^2,
\label{rcd3}
\eea
which has the full SU(3) gauge symmetry. This is because
the restricted potential, just as in SU(2), has the full gauge 
degrees of freedom.

With (\ref{cp3}) we have the Abelian decomposition of the SU(3)
gauge potential,
\bea
&\A_\mu=\hat A_\mu+\X_\mu
=\sum_p (\dfrac23 \hA_\mu^p+\W_\mu^p), \nn\\
&\X_\mu= \sum_p \W_\mu^p,  \nn\\
&\W_\mu^1= X_\mu^1 \n_1+ X_\mu^2 \n_2,
~~~\W_\mu^2=X_\mu^6 \n_6 + X_\mu^7 \n_7,  \nn\\
&\W_\mu^3= X_\mu^4 \n_4-  X_\mu^5 \n_5.
\label{cdec3}
\eea
Here again $\X_\mu$ transforms covariantly, and can be decomposed
to the three valence gluons $\W_\mu^p$ of the SU(2) subgroups. But
unlike $\hA_\mu^p$, they are mutually independent. So we have two 
binding gluons and six (or three complex) valence gluons in SU(3) 
QCD.

From (\ref{cdec3})  we have
\bea
&\hD _\mu \X_\nu=\sum_p \hD_\mu^p \W_\nu^p,
~~~\hD_\mu^p=\pro_\mu+ g \hA_\mu^p \times,   \nn\\
&\X_\mu\times \X_\nu=\sum_{p,q} \W_\mu^p \times \W_\nu^q,  \nn\\
&\vec{F}_{\mu\nu}=\hF_\mn + \hD _\mu \X_\nu -
\hD_\nu \X_\mu + g\X_\mu \times \X_\nu  \nn\\
&=\sum_p \big[\dfrac23 \hF_\mn^p
+ (\hD_\mu^p \W_\nu^p-\hD_\mu^p \W_\nu^p) \big]  \nn\\
&+\sum_{p,q}\W_\mu^p \times \W_\nu^q,
\eea
so that we have the following form of SU(3) ECD \cite{prd13,ijmpa14}
\bea
&{\cal L}=-\dfrac{1}{4}\hF_\mn^2
-\dfrac{1}{4}(\hD_\mu\X_\nu-\hD_\nu\X_\mu)^2 \nn\\
&-\dfrac{g}{2} (\hD_\mu \X_\nu
-\hD_\nu \X_\mu) \cdot (\X_\mu \times \X_\nu)\nn\\
&-\dfrac{g}{2} {\hF}_\mn \cdot (\X_\mu \times \X_\nu)
-\dfrac{g^2}{4} (\X_\mu \times \X_\nu)^2 \nn\\
&= \sum_p \Big\{-\dfrac{1}{6} (\hF_\mn^p)^2
-\dfrac{1}{4} (\hD_\mu^p \W_\nu^p- \hD_\nu^p \W_\mu^p)^2 \nn\\
&-\dfrac{g}{2} \hF_\mn^p \cdot (\W_\mu^p \times \W_\nu^p) \Big\} 
-\sum_{p,q} \dfrac{g^2}{4} (\W_\mu^p \times \W_\mu^q)^2 \nn\\
&-\sum_{p,q,r} \dfrac{g}2 (\hD_\mu^p \W_\nu^p- \hD_\nu^p \W_\mu^p)
\cdot (\W_\mu^q \times \W_\mu^r)  \nn\\
&-\sum_{p\ne q} \dfrac{g^2}{4} \big((\W_\mu^p \times \W_\nu^q)
\cdot (\W_\mu^q \times \W_\nu^p)  \nn\\
&+(\W_\mu^p \times \W_\nu^p)\cdot (\W_\mu^q \times \W_\nu^q) \big).
\label{3ecd}
\eea
This shows that the interactions in SU(3) QCD is more 
complicated than the SU(2) QCD. But what is remarkable 
about (\ref{3ecd}) is that it is Weyl symmetric, symmetric 
under the permutationof the three SU(2) subgroups of SU(3).

\begin{figure}
\begin{center}
\includegraphics[scale=0.5]{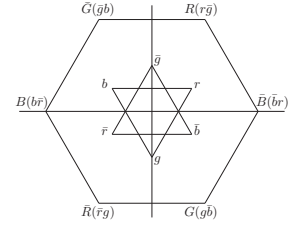}
\caption{\label{color} The color assignment of quarks and
chromons. The x-axis and y-axis represent the $\lambda_3$ and
$\lambda_8$ quantum numbers.}
\end{center}
\end{figure}

We can easily add  quarks in the Abelian decomposition,
\bea
&{\cal L}_{q} =\sum_k\bar \Psi_k (i\gamma^\mu D_\mu-m) \Psi_k \nn\\
&= \sum_k \Big[\bar \Psi_k (i\gamma^\mu \hD_\mu-m) \Psi_k
+\dfrac{g}{2} \X_\mu
\cdot \bar \Psi_k (\gamma^\mu \vec t) \Psi_k \Big] \nn\\
&=\sum_{p,k} \Big[\bar \Psi_k^p (i\gamma^\mu \hD_\mu^p-m) \Psi_k^p
+\dfrac{g}{2} \W_\mu^p \cdot \bar \Psi_k^p
(\gamma^\mu \vec \tau^p) \Psi_k^p \Big], \nn\\
&\hD_\mu = \pro_\mu + \dfrac{g}{2i} {\vec t}\cdot \hA_\mu,
~~\hD_\mu^p=\pro_\mu+\dfrac{g}{2i} {\vec \tau^p}\cdot \hA_\mu^p,
\label{qlag}
\eea
where $m$ is the mass, $k$ and $p$ denote the flavor and 
color of the quarks, and $\Psi_k^p$ represents the three
SU(2) quark doublets ({i.e., (r,b), (b,g), and (g,r) doublets)
of the (r,b,g) quark triplet.

From this it becomes obvious that the binding gluons and 
the valence gluons play different roles. So from now on we 
will call the binding gluon the ``neuron" ( or ``neuton") 
and the valence gluon the ``chromon" (or ``coloron").

To assign the color to the chromons, let $(r,g,b)$ be
the colors of three quarks. Then the colors of the six 
chromons are given by $(r\bb,b\bg,g\br,\br b,\bb g,\bg r)$, 
which we denote for simplicity by $(R,B,G,\bR,\bB,\bG)$. 
This is schematically shown in Fig.~\ref{color}.

We can show how the Abelian decomposition refines QCD 
interaction graphically. This is shown in Fig.\ref{ecdint}. 
In (A) the three-point QCD gluon vertex is decomposed to 
two vertices made of one neuron and two chromons and 
three chromons. In (B) the four-point gluon vertex is decomposed 
to three vertices made of one neuron and three chromons, 
two neurons and two chromons, and four chromons. In (C) 
the quark-gluon vertex is decomposed to the quark-neuron 
vertex and quark-chromon vertex. 

Notice that here (and in the following figures) the neurons are 
expressed by the wiggly lines (Maxwell part). This is because 
the monopole potential (Dirac part) makes the condensation, 
so that in the perturbative regime (inside the hadrons) it does 
not contribute to the Feynman diagrams. Also here three-point 
vertex made of three neurons or two neurons and one chromon, 
and four-point vertex made of three or four neurons are forbidden 
by the conservation of color. Moreover, the quark-neuron interaction 
does not change the quark color, but the quark-chromon interaction 
changes the quark color.

\begin{figure}
\begin{center}
\includegraphics[scale=0.5]{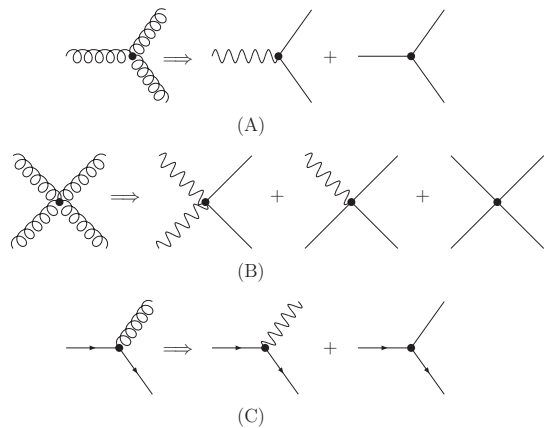}
\caption{\label{ecdint} The decomposition of vertices in SU(3) QCD. 
The three and four point gluon vertices are decomposed in (A) and
(B), and the quark gluon verteces are decomposed in (C). Notice that
here (and in the followings) the neurons are representedby wiggly 
lines and the chromons are represented by straight lines.}
\end{center}
\end{figure}

An important implication of Fig. \ref{ecdint} is that there are two 
types of gluon jets, the neuron jet and chromon jet. In principle 
we can test this experimentally by studying the gluon jets. 
Experiments can tell the difference between the photon-quark 
jet from the gluon-quark jet. If so, by (re-)analyzing the gluon-gluon 
jets and/or gluon-quark jets more carefully we could confirm that 
indeed there are two types of gluon jets. Experimental confirmation 
of this is very important, because this could endorse the existence 
of two types of gluons.

Our analysis tells that, although the Abelian decomposition
does not change QCD, it makes many hidden structures of 
QCD explicit. First, it tells that RCD is responsible for the 
confinement, because the valence gluons (being colored) 
have to be confined \cite{thooft,prd00}. So it makes the Abelian 
dominance obvious.

Second, it allows us to prove that the monopole is responsible 
for the area law in the Wilson loop integral. Indeed implementing 
(\ref{3cp}) on lattice, two lattice QCD groups (the SNU-KU and 
KEK-CU groups) independently performed a truly gauge independent 
lattice calculation, and showed that the monopole is responsible 
for the confinement \cite{kondo1,kondo2,cundy1,cundy2}. The SNU-KU 
result is shown in Fig. \ref{cundy}, which shows that the 
full gauge potential, the restricted potential, and the monopole 
potential all produce the linear confining potential in Wilson 
loop integral. This assures that we only need the monopole potential 
for the confining force.

\begin{figure}
\begin{center}
\includegraphics[scale=0.6]{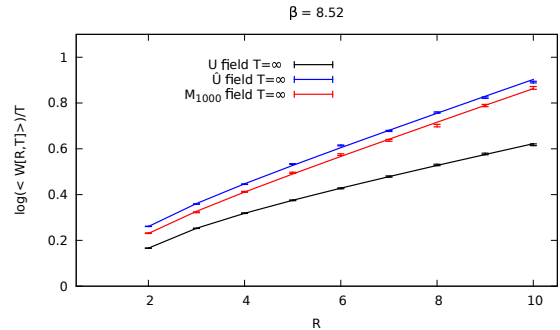}
\caption{\label{cundy} The lattice QCD calculation which establishes 
the monopole dominance in Wilson loop. Here the black, red, and 
blue slopes are obtained with the full potential, the restricted potential, 
and the monopole potential, respectively.}
\end{center}
\end{figure}

Moreover, the Abelian decomposition enlarges and doubles 
the gauge symmetry to the classical and the quantum 
gauge symmetries, because it automatically puts QCD 
in the background field formalism \cite{dewitt,prd01}. 
So the neurons and the chromons have independent gauge 
freedoms. This keeps both neurons and chromons massless.

Third, it reduces the complicated non-Abelian gauge 
symmetry to a simple discrete symmetry called the color
reflection invariance. To see this, consider the rotation of
basis called the color reflection in SU(2) QCD 
\bea
(\hn_1,\hn_2,\hn) \rightarrow (\hn_1,-\hn_2,-\hn).
\label{cref} 
\eea
Obviously this is a gauge transformation, so that this must 
remain a symmetry of QCD. On the other hand, the isometry 
condition (\ref{ap}) does not change under (\ref{cref}). This 
means that, after we select the Abelian direction $\n$ we have 
two different but gauge equivalent Abelian decompositions 
related by  the color reflection.  

What makes the color reflection symmetry so important is that 
it is the only remaining symmetry of the full gauge symmetry left 
over, after we make the Abelian decomposition \cite{prl81,prd81}. 
So the color reflection invariance plays the role of the non-Abelian 
gauge invariance after we have chosen the Abelian direction. This 
greatly simplifies us to implement the gauge invariance to calculate 
the QCD effective action \cite{prd13,ijmpa14}.

In the constant monopole background the effective action 
is given by
\bea
&{\cal L}_{eff} =- \sum_p \Big(\dfrac{H_p^2}{3}
+\dfrac{11g^2}{48\pi^2} H_p^2(\ln \dfrac{gH_p}{\mu^2}-c) \Big).
\label{effah}
\eea
where $H_p~(p=1,2,3)$ are the monopole background of three SU(2) 
subgroups. The corresponding effective potential has the true 
minimum at 
\bea
\langle H_1 \rangle=\langle H_2 \rangle=\langle H_3 \rangle
=\dfrac{\mu^2}{g} \exp \big(-\dfrac{16\pi^2}{11 g^2}
+\dfrac{3}{4} \big).  
\label{mgap}
\eea
The effective potential is shown in Fig. \ref{su3ep}. This 
demonstrates the monopole condensation which generates the 
desired mass gap in QCD. 

\begin{figure}
\begin{center}
\includegraphics[height=5cm, width=7cm]{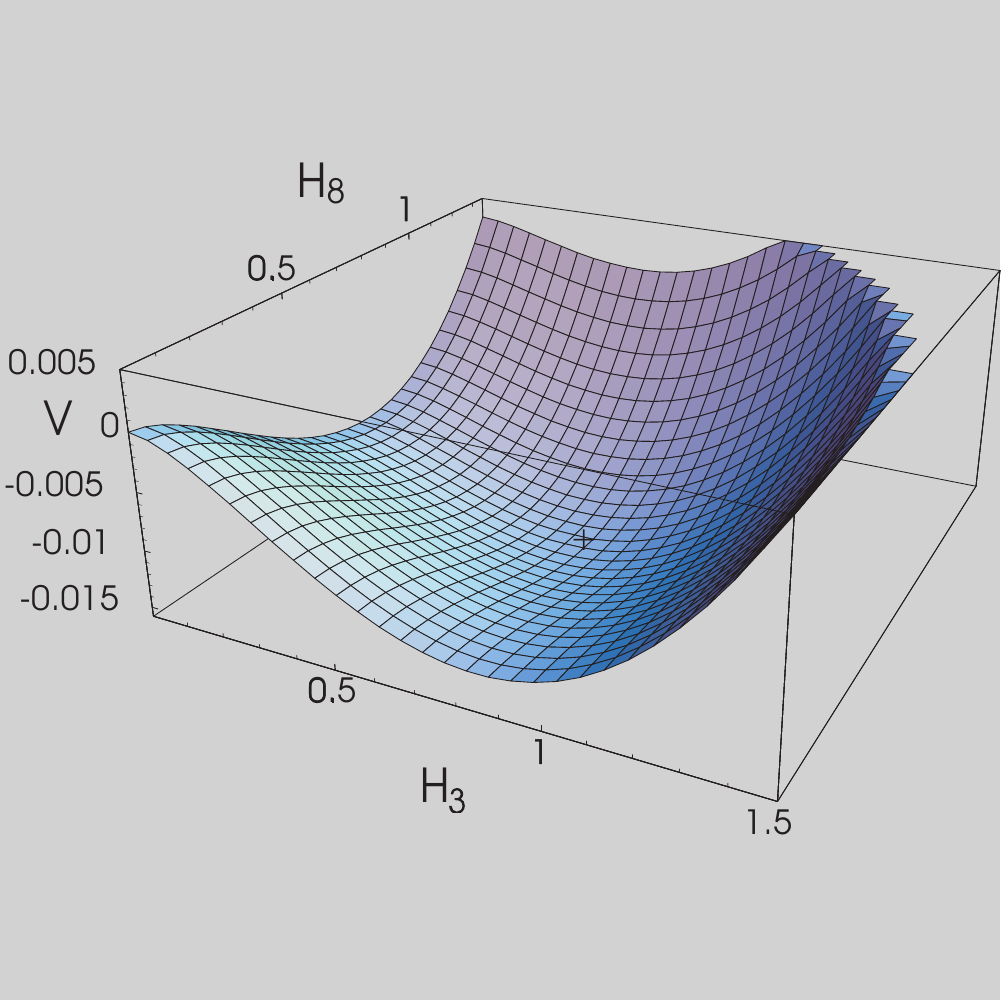}
\caption{\label{su3ep} The effective potential of SU(3) QCD.} 
\end{center}
\end{figure}

For the constant chromo-electric background we have the following
effective action 
\bea
&{\cal L}_{eff} =\sum_p \Big(\dfrac{E_p^2}{3}
+\dfrac{11g^2}{48\pi^2} E_p^2(\ln \dfrac{gE_p}{\mu^2}-c) \nn\\
&-i\dfrac{11g^2}{96\pi} E_p^2 \Big). \nn
\label{effae}
\eea
Notice that it has the imaginary part which is negative. 
This is very important, because this tells that the 
chromo-electric background induces the pair annihilation 
of  chromons \cite{prd13,ijmpa14,sch,prd02b,jhep05}. 

To summarize, the Abelian decomposition tells that QCD 
has two gluons which play different roles. Perturbatively 
(in terms of the Feynman diagrams) the neurons play 
the role of the photon and the chromons play the role of 
(massless) charged vector fields, in QED. Non-perturbatively, 
however, (the monopole part of) the neurons become 
the confining agents. In contrast, the chromons become 
the confined prisoners. Without the Abelian decomposition 
we can not tell this difference because all gluons are treated 
on equal footing.

\section{Glueballs and Oddballs}

The fact that the chromons become gauge covariant tells that 
they could form glueballs. For example, we can have the $g\bg$ 
or $ggg$ color singlet glueballs made of chromons which could 
be called the ``chromoballs". 

\begin{table}
\begin{center}
\caption{\label{gg} The possible quantum numbers for low-lying 
glueballs made of two chromons.  \\~\\}
\begin{tabular}{|c|c|c|c|c|c|}
\hline \hline
~~~$^{(2S+1)}L_J$~~~ & ~~~$J^{PC}$~~~ & possible candidates \\
\hline
$^1S_0$ ~~~ & $0^{++}$ & $f_0(500),f_0(980)$ \\
~~~~~~ $^5S_2$ & $2^{++}$ & $f_2(1950)$ \\
\hline
$^3P_0$ ~~~ & $0^{-+}$ & $\eta(1295),\eta(1405),\eta(1475)$  \\
$^3P_1$ ~~~ & $1^{-+}$ & ???  \\
$^3P_2$ ~~~ & $2^{-+}$ & $\eta_2(1645)$ \\
\hline
$^1D_2$ ~~~ & $2^{++}$ & Regge recurrence of $^1S_0$  \\
~~~~~~~ $^5D_0$ & $0^{++}$ & $f_0(1500)$  \\
~~~~~~~ $^5D_1$ & $1^{-+}$  & ???   \\
~~~~~~~ $^5D_2$ & $2^{++}$ & Regge recurrence of $^5S_2$ \\
\hline \hline
\end{tabular}
\end{center}
\end{table} 

Since we have six gauge covariant chromons ($R_\mu, B_\mu, G_\mu,
\bR_\mu,\bB_\mu, \bG_\mu$), we can construct low-lying color 
singlet glueballs with two ($g\bar g$) chromons whose wave functions 
are symmetric under the exchange 
\bea
&|g\bar g\rangle=\dfrac{|R_\mu \bR_\nu \rangle
+|B_\mu \bB_\nu \rangle +|G_\mu \bG_\nu \rangle}{\sqrt 3}.
\eea
The low-lying $g\bg$ glueball states classified by $^{(2S+1)}L_J$
are shown in Table \ref{gg}. In the table we have listed the 
possible candidates of the glueballs based on the PDG data, but
we emphasize that they are by no means certain.

Actually the number of the glueball states depends on 
how many degrees the chromons have. If we assume
the chromons to be massless they have only transversal 
degrees, but if we assume them massive they also have 
the longitudinal degrees. Here we have assumed that they 
acquire the (constituent) mass after the confinement sets 
in. But ultimately experiments should determine how many
degrees the chromons have. 

Similarly we can construct low-lying color singlet glueballs with 
three chromons,
\bea
&|ggg\rangle_d=\dfrac{\sum_{(RGB)}
|R_\mu B_\nu G_\rho \rangle}{\sqrt 6},  \nn\\
&|ggg\rangle_f=\dfrac{\sum_{[RGB]}
|R_\mu B_\nu G_\rho\rangle}{\sqrt 6},
\eea
where the sums in $ggg$ are the totally symmetric (the $d$-product)
and the totally anti-symmetric (the $f$-product) combination of three 
colors.

With this one can figure out the possible $ggg$ chromoball states.
The exact enumeration of three chromon bound states depends 
on the binding potential, but in a simple shell model one can construct 
the low-lying $ggg$ chromoball states \cite{coyne}. This is shown 
in Table \ref{ggg}. In general the $ggg$ glueballs are expected to 
be heavier than the $gg$ glueballs, because they have more chromons. 

\begin{table*}
\begin{center}
\caption{\label{ggg} The possible quantum numbers for low-lying
glueballs made of three chromons. Here S, A, and M mean symmetric, 
anti-symmetric, and mixed symmetries, and d and f mean (totally 
symmetric) d-product and (totally anti-symmetric) f-product of 
three chromons. \\~\\}
\begin{tabular}{|c|c|c|c|c|c|c|}
\hline \hline
Configuration & ~space~  & ~~spin~~ & ~color~ & ~~~$L$~~~ & 
~~~$S$~~~ & $J^{PC}$ \\
\hline
$(1s)^3$ & S & S & d & 0 & 1,3 & $1^{--},~3^{--}$ \\
         & S & A & f & 0 & 0 & $0^{-+}$ \\
\hline
$(1s)^2(1p)$ & S & S & d & 1 & 1,3 & ~~$(0,1,2)^{+-},(2,3,4)^{+-}$~~ \\
         & S & A & f & 1 & 0 & $1^{++}$ \\
         & M & M & d & 1 & 1,2 & $(0,1,2)^{+-},(1,2,3)^{+-}$ \\
         & M & M & f & 1 & 1,2 & $(0,1,2)^{++},(1,2,3)^{++}$ \\
\hline
$(1s)(1p)^2$ & S & S & d & $0,2$ & $1,3$ & $(1,2,3,4,5)^{--}$ \\
             & S & A & f & $0,2$ & $0$ & $0^{-+},2^{-+}$ \\
             & M & M & d & $0,2$ & $1,2$ & $(0,1,2,3,4)^{--}$ \\
             & M & M & f & $0,2$ & $1,2$ & $(0,1,2,3,4)^{-+}$ \\
             & M & M & d & $1$   & $1,2$ & $(0,1,2,3)^{--}$ \\
             & M & M & f & $1$   & $1,2$ & $(0,1,2,3)^{-+}$ \\
             & A & S & d & $1$   & $1,3$ & $(0,1,2,3,4)^{-+}$ \\
             & A & A & f & $1$   & $0$   & $1^{--}$ \\
\hline
$(1s)^2(2s)$ & S & S & d & $0$ & $1,3$ & $1^{--},3^{--}$ \\
             & S & A & f & $0$ & $0$ & $0^{-+}$ \\
             & M & M & d & $0$ & $1,2$ & $(1,2)^{--}$ \\
             & M & M & f & $0$ & $1,2$ & $(1,2)^{-+}$ \\
\hline
$(1s)^2(1d)$ & S & S & d & $2$ & $1,3$ & $(1,2,3,4,5)^{--}$ \\
             & S & A & f & $2$ & $0$ & $2^{-+}$ \\
             & M & M & f & $2$ & $1,2$ & $(0,1,2,3,4)^{--}$ \\
             & M & M & f & $2$ & $1,2$ & $(0,1,2,3,4)^{-+}$ \\
\hline \hline
\end{tabular}
\end{center}
\end{table*}

At this point one might wonder if the neurons could also form 
bound states. Certainly from the group theoretic point of view 
we could construct color singlet states with two or three neurons. 
This, however, does not guarantee that dynamically the neurons 
can make bound states. Since they carry no color charge the 
interaction among them should be very weak, so that they
are not likely to form bound states.

To clarify this point, consider the photons in QED. Clearly they 
interact among themself through the electron loops, but obviously 
they do not form bound states. Here the situation is very similar, 
because the neurons in QCD are exactly like the photons in QED. 
To amplify this point we show the possible interactions among 
neurons in Fig. \ref{nball}. This is precisely the photon interaction
of QED made of the charged vector field. 

From this we may conclude that the neurons do not make bound 
states. Indeed the Feynman diagram tells that, if such a bound 
state exists at all in QCD, it could be interpreted as a bound 
state of two quarkoniums. 

This should be compared with the possible Feynman diagram 
of the chromoball interactions shown in Fig. \ref{cball}. 
The contrast between the two Feynman diagrams are unmistakable. 
We emphasize that, without the Abelian decomposition, it would 
have been very difficult to see this difference. 

The above analysis tells that there must be a large number of 
glueballs. But experimentally we do not have many candidates 
of them. As we have remarked, one reason is that these glueballs 
may not exist as mass eigenstates, because they could mix 
with $q\bq$ states. So it is very important to discuss the 
glueball-quarkonium mixing to identify these glueballs.

Another reason is that the chromoballs (unlike the quarkoniums) 
have an intrinsic instability, because they tend to annihilate 
each other in strong chromo-electric field \cite{prd13,ijmpa14}. 
This is due to the anti-screeing and asymptotic freedom \cite{wil,pol}. 

\begin{figure}
\begin{center}
\includegraphics[scale=0.6]{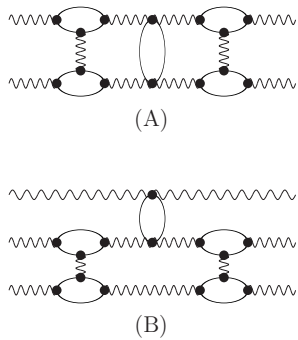}
\caption{\label{nball} The possible Feynman diagrams of the
neuron interaction.}
\end{center}
\end{figure}

We can estimate the glueball partial decay width coming 
from this instability. According to the QCD one-loop 
effective action (\ref{effae}) the chromon annihilation 
probability per unit voloume per unit time is given by 
\bea
\Gamma_A=\sum_p \dfrac{11g^2}{96\pi} {\bar E}_p^2 
\times \dfrac{4\pi}{3\Lambda_{QCD}^3},
\eea
where the sum is on three SU(2) subgroups and ${\bar E}_p$ 
is the average chromo-electric field of each subgroup inside 
the glueballs. Now, if we choose $\alpha_s \simeq 0.4$, 
$\Lambda_{QCD}\simeq 339~MeV$ (for three quark flavors), 
and $\bar E_p \simeq (g/\pi)\Lambda^2_{QCD}$ we have 
$\Gamma_A \simeq 398~MeV$ \cite{pdg}. But notice that 
with $\Lambda_{QCD} \simeq 200~MeV$, we have
$\Gamma_A\simeq 235~MeV$ \cite{pesk}.

Of course this is a rough estimate, but notice that this is the 
partial decay width we expect from the asymptotic freedom, 
in addition to the ``normal" hadronic decay width. This strongly 
implies that in general the glueballs (in particular excited 
ones) are expected to be quite unstable. As we have remarked 
this could be one of the reasons why there are so few candidates 
of glueballs experimentally.  

\begin{table}[htbp]
\begin{center}
\caption{\label{qn} The possible quantum numbers for
low-lying the $q\bq$, $g\bg$, and $ggg$ states.
\\~\\}
\begin{tabular}{|c|c|c|c||c|c|c|c|}
\hline \hline
State &~~$q\bq$~~&~~$g\bg$~~&~~$ggg$~~& 
State &~~$q\bq$~~&~~$g\bg$~~&~~$ggg$~~  \\
\hline \hline
State &~~$q\bq$~~&~~$g\bg$~~&~~$ggg$~~& 
State &~~$q\bq$~~&~~$g\bg$~~&~~$ggg$~~  \\
\hline
$0^{++}$ & O & O & O & $2^{++}$ & O & O & O \\
\hline
$0^{+-}$ & X & X & O & $2^{+-}$ & X & X & O \\
\hline
$0^{-+}$ & O & O & O & $2^{-+}$ & O & O & O \\
\hline
$0^{--}$ & X & X & O & $2^{--}$ & O & X & O \\
\hline
$1^{++}$ & O & O & O & $3^{++}$ & O & O & O \\
\hline
$1^{+-}$ & O & X & O & $3^{+-}$ & O & X & O \\
\hline
$1^{-+}$ & X & O & O & $3^{-+}$ & X & O & O \\
\hline
$1^{--}$ & O & X & O & $3^{--}$ & O & X & O \\
\hline \hline
\end{tabular}
\end{center}
\end{table}

Although the glueballs in general mix with the quarkoniums, 
in particular cases the pure glueballs could exist \cite{coyne}. 
This is because some of the $g\bg$ glueballs have the quantum 
number $J^{PC}$ which $q\bar q$ can not have. In the quark 
model the $q\bq$ states in the natural spin-parity series 
$P=(-1)^J$ must have spin one, and hence $CP=+1$. So 
the mesons with natural spin-parity and $CP=-1$ (e.g., 
$0^{--}, 0^{+-}, 1^{-+}, 2^{+-}$, etc.) are forbidden. But 
the $g\bg$ or $ggg$ glueballs could have these quantum 
states. In fact, the $ggg$ glueballs, unlike $q\bq$, could have 
all possible $J^{PC}$.

So these particular glueballs carrying the quantum numbers
which $q\bq$ can not have can not mix with the quarkoniums,
and they are called the ``oddballs" \cite{coyne}. The 
low-lying oddballs become important because they could be 
observed as pure glueball states. The Table \ref{qn} summarizes 
the possible $J^{PC}$ for the $q\bq$, $g\bg$, and $ggg$ states. 
From this we can say definitely that any of low-lying $0^{+-},
~0^{--},~1^{-+}$, or $2^{+-}$ meson states must be pure 
glueballs. This could provide a crucial information for us to 
search for the pure glueballs.

\begin{figure}
\includegraphics[scale=0.6]{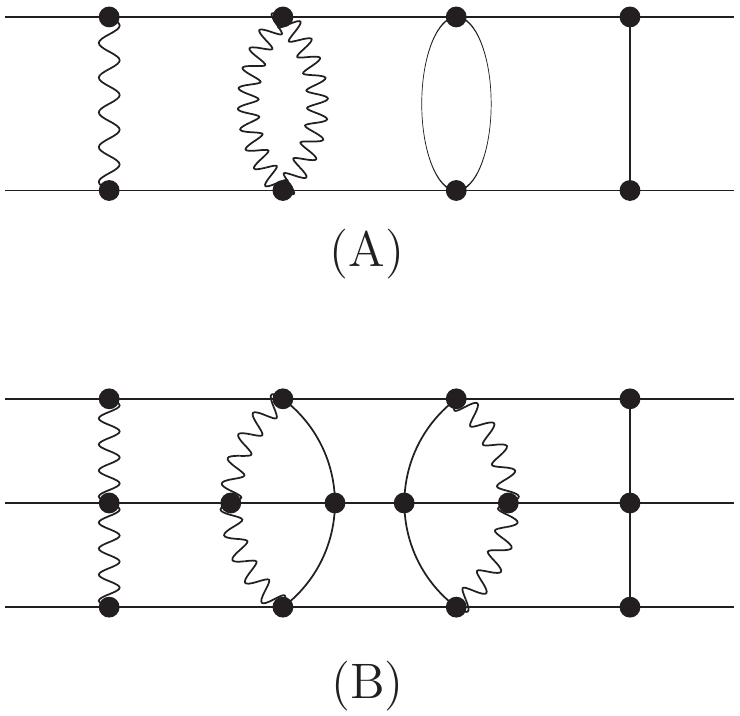}
\caption{\label{cball} The possible Feynman diagrams which 
bind the chromons. Two chromon binding is shown in (A), three
chromon binding is shown in (B).}
\end{figure}

The above analysis tells that the identification of the glueballs
may not be simple. To identify these glueball states we have to 
compare the theoretical prediction with  the experimental data. 
In the Appendix we show two tables, the one which provides 
the standard quark model interpretation of the low-lying mesons 
and the other which contains the light iso-singlet mesons 
which can not easily be identified as $q\bq$ states, from 
PDG data \cite{pdg}.

\section{Glueball-Quarkonium Mixing}

To identify the glueballs we have to study their mixing with 
the quarkoniums. But before we discuss the mixing, it is worth 
discussing the $q\bar{q}$ octet-singlet mixing in the quark 
model first. 

The $q\bar{q}$ binding energy may come from two orthogonal 
processes, the exchange and annihilation processes. Let us 
assume \cite{snu89}
\bea
&\langle u\bar{u}|H|u\bar{u} \rangle_{Ex}
=\langle d\bar{d}|H|d\bar{d} \rangle_{Ex}=E, \nn \\
&\langle s\bar{s}|H|s\bar{s} \rangle_{Ex}=E'=E+\Delta,    \nn  \\
&\langle q'\bar{q'}|H|q\bar{q} \rangle_{An}=A,
~~~(\rm{for~all~q,q'}).
\eea
Now with
\bea
&|8 \rangle=\dfrac{|u\bar u\rangle+|d\bar d \rangle
-2|s\bar s\rangle}{\sqrt 6}, \nn\\
&|1 \rangle=\dfrac{|u\bar u \rangle+|d\bar d \rangle
+|s\bar s \rangle}{\sqrt 3},
\eea
we may obtain the following mass matrix for the $q\bar{q}$
which describes the octet-singlet mixing,
\bea
&M^2=\left( \begin{array}{cc}  \langle 8 |H|8  \rangle
 & \langle 8 |H| 1  \rangle  \\
 \langle 1 |H| 8  \rangle   &   \langle 1 |H| 1  \rangle
\end{array} \right) \nn\\
&=\left( \begin{array}{cc}
E+\dfrac{2}{3}\Delta & -\dfrac{\sqrt{2}}{3}\Delta \\
-\dfrac{\sqrt{2}}{3}\Delta  & E+\dfrac{1}{3}\Delta +3A
\end{array} \right).
\eea
Notice that $\Delta$-term is responsible for the mixing.

From this we have the mass eigenvalues
\bea
& m^2_\pm= \dfrac12 \big[(E'+E+3A) \pm D \big], \nn\\
&D=\sqrt{(E'-E-A)^2+8A^2}.
\eea
Notice that (when A is positive) the eigenvalues $m^2_{\pm}$
must satisfy $m^2_-< E$ and $E' < m^2_+$. This tells that
the annihilation contribution has a tendency to make the mass
splitting larger, which seems to be the case in reality.

Now we can discuss the glueball-quarkonium mixing. The 
possible Feynman diagrams for the mixing is shown in 
Fig. {\ref{mixing}. From this it is clear that the mixing takes 
place not just between the quarkoniums and glueballs but 
also between the $gg$ and $ggg$ glueballs, directly or 
through the virtual states made of neurons. 

\begin{figure}
\begin{center}
\includegraphics[scale=0.6]{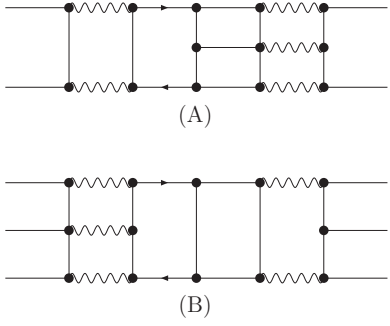}
\caption{\label{mixing} The possible glueball-quarkonium mixing
diagrams.}
\end{center}
\end{figure}

To proceed, let $|G\rangle$ be the glueball state which mixes with 
two quarkonium states $|8\rangle, |1\rangle$ and consider the mass
matrix of $(|8\rangle, |1\rangle, |G\rangle)$,
\bea
&{\cal M}= \left(  \begin{array}{ccc}
a  &  b  & 0\\
b  & c  & d \\
0  &  d & e \end{array} \right),
\eea
whose eigenvalues are given by $\lambda_i$. In this case the
mixing matrix $\cal U$ which transforms the unphysical states
to the physical states $(|m_1\rangle, |m_2\rangle, |m_3\rangle)$
and diagonalizes $\cal M$ to $\cal D$ is given by \cite{snu89}
\bea
&{\cal D}={\cal U M U^\dagger}
={\rm diag}~(\lambda_1, \lambda_2, \lambda_3), \nn\\
&{\cal U}=\left( \begin{array}{ccc}
\dfrac{b(\lambda_1-e)}{d(\lambda_1-a)}\alpha_1,
&\dfrac{\lambda_1-e}{d} \alpha_1, & \alpha_1 \\
\dfrac{b(\lambda_2-e)}{d(\lambda_2-a)}\alpha_2,
& \dfrac{\lambda_2-e}{d} \alpha_2, & \alpha_2 \\
\dfrac{b(\lambda_3-e)}{d(\lambda_3-a)} \alpha_3,
& \dfrac{\lambda_3-e}{d} \alpha_3 & \alpha_3
\end{array} \right), \nn\\
& \alpha_i=\dfrac{d}{(\lambda_i-e)
\sqrt {1+(\dfrac{b}{\lambda_i-a})^2
+(\dfrac{d}{\lambda_i-e})^2}}.
\label{mmatrix}
\eea
Moreover, we have the sum rules
\bea
&a+c+e=\lambda_1+ \lambda_2 +\lambda_3,  \nn\\
&ac+ce+e a-b^2-d^2=\lambda_1 \lambda_2
+ \lambda_2 \lambda_3 +\lambda_3 \lambda_1,  \nn\\
&a c e-b^2 e-d^2 a=\lambda_1 \lambda_2 \lambda_3.
\eea
Notice that $\alpha_i$ determines the gluon content of
physical states.

The gluon content of the physical states has important implication. 
For example, this allows us to predict the relative branching ratios 
of $\psi$ to $\gamma X$ decays among the physical states. This
is because the $\psi$ decay process to ordinary non-charming
physical states is the Okubo-Zweig-Iizuka (OZI) suppressed
process which can only be made possible through the gluons. 

So, except for the kinematic phase factor, the glue content of the
physical states determines the radiative decay branching ratios.
This means that for the S wave decay (i.e., for $0^{++}$ and 
$2^{++}$)
we have
\bea
&R\Big(\dfrac{\psi \rightarrow \gamma X_k}{\psi\rightarrow 
\gamma X_i}\Big)=\Big(\dfrac{\alpha_k}{\alpha_i}\Big)^2
\Big(\dfrac{m^2_\psi-m^2_k}{m^2_\psi-m^2_i} \Big)^3,
\eea
but for the P wave decay (i.e., for $0^{-+}$) we expect to have
\bea
&R\Big(\dfrac{\psi \rightarrow \gamma X_k}{\psi\rightarrow 
\gamma X_i}\Big)=\Big(\dfrac{\alpha_k}{\alpha_i}\Big)^2
\Big(\dfrac{m^2_\psi-m^2_k}{m^2_\psi-m^2_i} \Big)^5,
\eea
where the last term is the kinematic phase space factor. So 
the gluon content of the physical states can explain the underlying
dynamics of the OZI rule.

Of course the idea of the glueball-quarkonium mixing has been 
suggested many times before \cite{coyne,ams2,close}. But the 
clear picture of the mixing was lacking because the constituent 
gluons were not well-defined. The quark and chromon model allows 
us to discuss the mixing without any ambiguity \cite{snu89}. 

\begin{table*}[htbp]
\label{t0++}
\caption{The predicted mass of the third physical state, 
the quark and glue component (the probability) of the 
physical states, and the relative radiative decay ratios 
for fixed values of the gluon mass $\mu$ in the $0^{++}$ 
channel. Here we choose $f_0(1500)$ and $f_0(1710)$ as the input.}
\begin{center}
\begin{tabular}{cccccccccccccccc}
\hline
\hline
 $\mu$&$A$&$\nu$&$\zeta$&$m_3$ & \multicolumn{3}{c} {$m_1=f_0(1500)$} & \multicolumn{3}{c} {$m_2=f_0(1710)$} & \multicolumn{3}{c} {$m_3$} & $R(m_2/m_1)$ & $R(m_3/m_1)$\\
$~$&~&~&~&~& $u+d$ & $s$ & $g$ & $u+d$ & $s$ & $g$ & $u+d$ & $s$ & $g$&~&~\\
\hline
$0.76$& 0.27  & 0.18 & 0.12  & 1.40 & 0.07 & 0.00 & 0.93 & 0.73 & 0.20 & 0.07 & 0.19 & 0.80 & 0.00& 0.00 & 0.05\\
\hline
$0.78$& 0.23  & 0.31 & 0.42  & 1.40 & 0.26 & 0.01 & 0.73 & 0.59 & 0.16 & 0.25 & 0.15 & 0.83 & 0.02& 0.02 & 0.14\\
\hline
$0.80$& 0.18 & 0.36 & 0.69  & 1.39 & 0.44 & 0.01 & 0.54 & 0.45 & 0.12 & 0.43 & 0.11 & 0.87 & 0.02& 0.05 & 0.59\\
\hline
$0.82$& 0.14 & 0.35 & 0.90  & 1.39 & 0.62 & 0.02 & 0.36 & 0.30 & 0.08 & 0.62 & 0.09 & 0.90 & 0.01& 0.07 & 1.26\\
\hline
$0.84$& 0.09 & 0.29 & 0.92  & 1.39 & 0.79 & 0.02 & 0.18 & 0.15 & 0.04 & 0.80 & 0.05 & 0.93 & 0.01& 0.09 & 3.26\\
\hline
$0.86$& 0.04 & 0.07 & 0.12  & 1.39 & 0.96 & 0.03 & 0.01 & 0.01 & 0.00 & 0.99 & 0.03 & 0.97 & 0.00& 0.12 & 85.71\\
\hline
\hline
\end{tabular}
\end{center}
\end{table*}

\section{Examples of Mixing: Numerical Analysis}

To discuss the mixing notice that, among the five low-lying 
$g\bg$ states ($^1S_0,~^5S_2,~^3P_0,~^3P_1,~^3P_2$ states) 
in Table I there are three glueball states (i.e., $0^{++},~2^{++},
~{\rm and} ~0^{-+}$ states) which can easily mix with the 
low-lying quarkonium states. So in this section we will restrict 
ourselves to the mixing of these glueball states with the 
corresponding isosinglet $q\bq$ states below 2GeV---with 
the light quarks $u$, $d$, and $s$ only---for simplicity.

In this approximation the mass matrix of the mixing can be 
written as  
\bea
M^2  =\left(  \begin{array}{ccc}
E+\dfrac{2}{3}\Delta  & -\dfrac{\sqrt{2}}{3}\Delta  & 0\\
-\dfrac{\sqrt{2}}{3}\Delta  & E+\dfrac{1}{3}\Delta +3A & \nu\\
0  &  \nu & G \end{array} \right).
\label{massm}
\eea
It has five parameters, but we can fix $E$ and $\Delta$ from 
the $q\bq$ octet-singlet mixing. So we need three inputs to 
fix the mass matrix completely. 

There are different ways to fix the mass matrix. One way 
is to choose two predominantly $q\bq$ states, or simply 
to choose two lowest mass eigenstates, from PDG. With this 
we could treat $G$ as a free parameter, and find (if possible) 
the best fit for $G$ which could explain the PDG data. In 
this case we can replace $G$ with the chromon constituent 
mass $\mu$ writing $G=4 \mu^2$, since $G$ represents the 
mass of two chromons.  

Another way to fix the mass matrix is to notice that in this
approximation we may assume
\bea
\langle q'\bq'|H|q\bq \rangle_{An} \simeq 
\langle q'\bq'|gg \rangle \langle gg|q\bq \rangle.
\eea
So, in stead of varying $\mu$ we could impose
the condition $3A=\nu^2$ to fix the mass matrix.
But this requirement could be too stringent, and we 
will not require this in this paper.

We emphasize the clarity of our mixing mechanism presented
by the quark and chromon model. All terms in (\ref{massm})
have clear physical meaning. For example we can draw the 
Feynman diagram which represents the isosinglet-glueball 
mixing parameter $\nu$, and could in principle calculate it 
theoretically. 
 
With this we can predict the mass of the third state, calculate 
the quark and gluon contents of the physical states, and the 
relative branching ratios of the $\psi$ radiative decay to the 
physical states in each channel (in terms of $\mu$ if necsssary). 

With this strategy we now can discuss the glueball-quarkonium 
mixing in each channel separately. According to PDG the 
low-lying iso-singlet physical states in the $0^{++}$ and 
$2^{++}$ channels below 2 GeV are 
$f_0(500),~f_0(980),~f_0(1370),~f_0(1500),~f_0(1710)$ and  $f_2(1270),~f_2'(1525),~f_2(1950)$. In the $0^{-+}$ channel 
we have $\eta(548),~\eta'(958),~\eta(1295),~\eta(1405),
~\eta(1475)$ and $\eta(1760)$ \cite{pdg}. These are the subjects 
of our analysis in the following. 

\begin{table*}[htbp]
\label{t2++}
\caption{The numerical analysis of the mixing in the $2^{++}$ 
channel, with $f_2(1270)$ and $f_2(1950)$ as the input.}
\begin{center}
\begin{tabular}{cccccccccccccccc}
\hline
\hline
 $\mu$&$A$&$\nu$&$\zeta$&$m_3$ & \multicolumn{3}{c} {$m_1=f_2(1270)$} & \multicolumn{3}{c} {$m_2=f_2(1950)$} & \multicolumn{3}{c} {$m_3$} & $R(m_2/m_1)$ & $R(m_3/m_1)$\\
$~$&~&~&~&~& $u+d$ & $s$ & $g$ & $u+d$ & $s$ & $g$ & $u+d$ & $s$ & $g$&~&~\\
\hline
$0.76$& 0.39  & 0.95 & 2.33  & 1.47 & 0.40 & 0.00 & 0.60 & 0.35 & 0.36 & 0.29 & 0.25 & 0.64 & 0.11& 0.19 & 0.15\\
\hline
$0.78$& 0.35  & 0.99 & 2.78  & 1.47 & 0.46 & 0.01 & 0.53 & 0.33 & 0.33 & 0.34 & 0.22 & 0.66 & 0.12& 0.25 & 0.18\\
\hline
$0.80$& 0.31  & 1.01 & 3.26  & 1.48 & 0.52 & 0.01 & 0.47 & 0.30 & 0.30 & 0.40 & 0.18 & 0.69 & 0.12& 0.33 & 0.21\\
\hline
$0.82$& 0.28  & 1.02 & 3.79  & 1.48 & 0.58 & 0.01 & 0.41 & 0.27 & 0.27 & 0.46 & 0.15 & 0.72 & 0.13& 0.43 & 0.24\\
\hline
$0.84$& 0.24  & 1.02 & 4.38  & 1.49 & 0.64 & 0.01 & 0.36 & 0.24 & 0.24 & 0.52 & 0.13 & 0.75 & 0.12& 0.57 & 0.27\\
\hline
$0.86$& 0.20  & 0.99 & 5.06  & 1.49 & 0.69 & 0.01 & 0.30 & 0.20 & 0.21 & 0.59 & 0.10 & 0.78 & 0.11& 0.76 & 0.30\\
\hline
\hline
\end{tabular}
\end{center}
\end{table*}

\subsection{$0^{++}$ channel}
 
In this channel one would normally assume $a_0(980)$ to be 
the isotriplet partner of the isosinglet $q\bq$ and choose 
\bea
&E=a_0^2, ~~~~ a_0=a_0(980), \nn \\
&\Delta =2(K^{2}-a_0^2),~~~~K=K_0^*(1430).
\label{0++1}
\eea
This seems natural because $a_0(980)$ which is supposed 
to be made of $u$ and $d$ quarks is lighter than 
$K_0^*(1430)$ made of $s$ quark. 

On the other hand PDG interprets $a_0(980)$ (as well as 
$f_0(500)$ and $f_0(980)$) to be a meson-meson bound state, 
and suggests the following choice \cite{pdg}
\bea
&E=a_0^2, ~~~~ a_0=a_0(1450), \nn \\
&\Delta =2(K^{2}-a_0^2),~~~~K=K_0^*(1430).
\label{0++2}
\eea
This looks somewhat strange because this implies that the 
$q\bq$ state made of $u+d$ quark is heavier (or at least 
not lighter) than the $q\bq$ state made of the $s$ quark. 

Clearly the numerical analysis of the mixing will depend 
very much on which imput we use, and it is not clear which 
view is correct. But here we will simply adopt the PDG 
suggestion and use (\ref{0++2}) as the input in our analysis.

With this we have three undetermined parameters in the mass 
matrix. To fix them we may choose two physical states from 
PDG, and vary the chromon mass $\mu$ as an independent 
parameter. But here we have five physical states, $f_0(500),
~f_0(980),~f_0(1370),~f_0(1500)$, and $f_0(1710)$ below 
2 GeV. Since the identity of $f_0(500)$ and $f_0(980)$ are 
not clear we will choose $f_0(1500)$ and $f_0(1710)$ as 
the input. In this case we obtain Table \ref{t0++}. Notice that 
we have calculated $\zeta=\nu^2/A$ to see how good is 
the constraint $3A=\nu^2$ in this approximation.  

The numerical result suggests that the mass of the third state 
is around 1400 MeV which is predominantly a $s\bar s$ state, 
which we could interpret to be $f_0(1370)$. Interestingly, 
the physical contents of two other states depend very much 
on the value of the chromon mass $\mu$. When the mass 
is around 760 MeV, $f_0(1500)$ become predominantly the 
glue state. But as the chromon mass increases to 860 MeV, it 
becomes a $u+d$ state and $f_0(1710)$ quickly becomes 
the glue state. 

So when the chromon mass is around 760 MeV the above result 
appears to be in agreement with the suggestion of PDG, which 
lists $f_0(1370)$ and $f_0(1710)$ as the $q\bq$ states \cite{pdg}.
But here again the $q\bq$ state made of $s$ quark becomes 
lighter than the $q\bq$ state made of $u+d$. This, of course,
is due to the input (\ref{0++2}). 

In principler we could determine the chromon mass with 
our prediction of the relative ratio of the $\psi$ radiative 
decay. Unfortunately at the moment PDG has no experimental 
data available for us to do this. 

\subsection{$2^{++}$channel}

In this channel we have three physical states below 2 GeV, 
$f_2(1270)$, $f_2'(1525)$, and $f_2(1950)$. Of course, we 
also have $f_2(1430)$, $f_2(1565)$, $f_2(1640)$, $f_2(1810)$, 
and $f_2(1910)$, but we will not consider them here because
PDG does not classify them as established states. On the other 
hand the fact that there are so many candidates of $2^{++}$ 
states implies that we need more caution to analyse this 
channel.

Now, we can choose  
\bea
&E=a^2_2,~~~a_2=a_2(1320), \nn \\
&\Delta =2(K^{*2}-a^2_2),~~~~K^*=K^*(1430), 
\eea
as the input and vary the chromon mass $\mu$ as a free 
parameter. In this case we have three possibilities to choose 
two input states from $f_2(1270)$, $f_2'(1525)$, and 
$f_2(1950)$. 

With $f_2(1270)$ and $f_2(1950)$ as the input, we obtain 
Table \ref{t2++}. Notice that when $\mu\simeq 760$ MeV, 
we have $m_3\simeq 1,470$ MeV which could be identified 
as $f'(1525)$. In this case $f_2(1270)$ becomes a mixture 
of $u+d$ and glue states, and $f_2(1950)$ becomes a mixture 
of $u+d$, $s$ and glue states. But the third physical state 
$f'(1525)$ becomes predominantly an $s\bar s$ state. 

But when the chromon mass $\mu$  becomes around 860 MeV, 
$f_2(1270)$ becomes predominantly $u+d$ state and the third 
state $f_2'(1525)$ becomes predominantly $s\bs$ state. This 
is in line with the PDG suggestion, which interprets $f_2(1270)$ 
and $f_2'(1525)$ as the $q\bq$ states \cite{pdg}. 

\begin{table*}[htbp]
\label{t0-+}
\caption{The numerical amalysis of the mixing in the $0^{-+}$
channel. Here we choose $\eta'(958)$, $\eta(1405)$, and 
$\eta(1760)$ as the input.}
\begin{center}
\begin{tabular}{cccccccccccccccccccccc}
\hline
\hline
$\mu$ & $m_4$ & \multicolumn{4}{c} {$m_1=\eta'(958)$}
& \multicolumn{4}{c} {$m_2=\eta(1405)$} & \multicolumn{4}{c} {$m_3=\eta(1760)$}
& \multicolumn{4}{c} {$m_4$} \\
$~$&~& $u+d$ & $s$ & $2g$ &  $3g$ &$u+d$ & $s$ & $2g$&  $3g$ & $u+d$ & $s$ & $2g$&  $3g$ & $u+d$ & $s$ & $2g$&  $3g$ \\
\hline
$0.50$  &  0.55 & 0.02 & 0.03 & 0.93 & 0.02 & 0.13 & 0.11 & 0.05 & 0.72 & 0.43 & 0.30 & 0.01 & 0.26 & 0.43 & 0.57 & 0.00 & 0.00 \\
\hline
$0.50$  &  0.55 & 0.01 & 0.01 & 0.96 & 0.03 & 0.16 & 0.13 & 0.01 & 0.70 & 0.41 & 0.28 & 0.04 & 0.27 & 0.43 & 0.57 & 0.00 & 0.00 \\
\hline
$0.52$  &  0.54 & 0.04 & 0.07 & 0.85 & 0.04 & 0.20 & 0.17 & 0.13 & 0.50 & 0.31 & 0.22 & 0.01 & 0.46 & 0.45 & 0.54 & 0.00 & 0.00 \\
\hline
$0.52$  &  0.55 & 0.00 & 0.01 & 0.92 & 0.07 & 0.29 & 0.25 & 0.01 & 0.45 & 0.26 & 0.18 & 0.07 & 0.48 & 0.44 & 0.56 & 0.00 & 0.00 \\
\hline
$0.54$  &  0.54 & 0.06 & 0.12 & 0.76 & 0.06 & 0.26 & 0.22 & 0.23 & 0.28 & 0.20 & 0.14 & 0.00 & 0.66 & 0.47 & 0.52 & 0.01 & 0.00 \\
\hline
$0.54$  &  0.54 & 0.00 & 0.00 & 0.88 & 0.11 & 0.44 & 0.37 & 0.01 & 0.19 & 0.11 & 0.08 & 0.11 & 0.70 & 0.45 & 0.55 & 0.00 & 0.00 \\
\hline
\hline
\end{tabular}
\end{center}
\end{table*}

\begin{table*}[htbp]
\begin{center}
\begin{tabular}{ccccccccc}
\hline
\hline
 $\mu$ & $m_4$ & $R(m_2/m_1)$ & $R(m_3/m_1)$
 & $R(m_4/m_1)$ & A & $\nu$ & $\epsilon$ \\
\hline
$0.50$  &  0.55 & 0.12 & 0.06 & 0.004  & 0.84 & 0.34 & -0.07  \\
\hline
$0.50$  &  0.55 & 0.46 & 0.13 & 0.003  & 0.84 & 0.30 & 0.28  \\
\hline
$0.52$  &  0.54 & 0.03 & 0.08 & 0.006  & 0.75 & 0.40 & -0.13  \\
\hline
$0.52$  &  0.55 & 0.44 & 0.33 & 0.004  & 0.75 & 0.31 & 0.47  \\
\hline
$0.54$  &  0.54 & 0.00 & 0.10 & 0.009  & 0.66 & 0.42 & -0.20  \\
\hline
$0.54$  &  0.54 & 0.26 & 0.64 & 0.003  & 0.66 & 0.24 & 0.64  \\
\hline
\hline
\end{tabular}
\end{center}
\end{table*}

Experimentally, PDG shows 
\bea
&J/\Psi \rightarrow \gamma f_2(1270) 
\simeq (1.43\pm 0.11)\times 10^{-3}  \nn\\
&J/\Psi \rightarrow \gamma f_2'(1525) 
\simeq (4.5 +0.7-0.4) \times 10^{-4},  \nn
\eea
which implies
\bea
R\big(f_2'(1525)/f_2(1270)\big) \simeq 0.31 \pm 0.05.
\eea 
Remarkably this agrees excellently with our prediction in 
Table \ref{t2++}, when the chromon mass becomes 860 MeV. 
So all in all the mixing in this channel seems to work very well, 
although we certainly need a more careful analysis.

\subsection{$0^{-+}$ channel}

In this channel we have six physical states below 2 GeV, $\eta(548),~\eta'(958),~\eta(1295),~\eta(1405),~\eta(1475)$,
and $\eta(1760)$. But here we need a special attention 
because of the expected difficulties (the $U(1)$ problem, 
PCAC, etc.) in this channel. Moreover, Table II tells that 
there is $(1s)^3$ $ggg$ glueball state which can mix with 
the other states. 

So we generalize the mixing matrix to the $4\times 4$ matrix
\bea
M^2 =\left( \begin{array}{cccc}
E+\dfrac{2}{3}\Delta  & -\dfrac{\sqrt{2}}{3} \Delta & 0 & 0 \\
-\dfrac{\sqrt{2}}{3}\Delta  & E+\dfrac{1}{3}\Delta 
+3A & \nu & \nu'\\
0  &  \nu & G & \epsilon  \\
0 & \nu' & \epsilon & G'  \end{array} \right)
\eea
to include the $ggg$ state. This has eight parameters,
but we may express $G$ and $G'$ by the chromon mass 
$\mu$ and put $G=4 \mu^2$ and $G'=9 \mu^2$.
This reduces the number of the parameters to seven. 

Now, with
\bea
&E=\pi^2,~~~\pi=\pi(140), \nn \\
&\Delta =2(K^2-\pi^2),~~~~K=K(498),
\label{0-+}
\eea
as the input we have to fix five more parameters. To do 
that we may impose the condition $\nu'=3/2 \nu$, because 
$\nu$ and $\nu'$ represent two and three gluon couplings 
to the iso-singlet $q\bq$. With this we can choose three 
physical states as the input and vary the chromon mass 
$\mu$ to predict the mass of the fourth physical state. 

We could also try the condition $3A= \nu^2+ \nu'^2$,
assuming
\bea
&\langle q'\bq'|H|q\bq \rangle_{An} \simeq 
\langle q'\bq'|gg \rangle \langle gg|q\bq \rangle \nn\\
&+\langle q'\bq'|ggg \rangle \langle ggg|q\bq \rangle.
\eea
But again this constraint could be too stringent.

Now, if we choose $\eta'(958),~\eta(1405)$ and $\eta(1760)$ 
as the input, we obtain Table VI. Notice that here we have 
two sets of solution, because the $4 \times 4$ mixing involves
quadratic equation.
 
In this analysis the mass of the fourth physical state becomes 
around 550 MeV, which could be interpreted to be $\eta(548)$. 
The result shows that the physical contents of $\eta(1405)$ 
and $\eta(1760)$ depend very much on the mass of the gluon. 
On the other hand here $\eta(548)$ is a mixture of $u+d$ and 
$s$, but $\eta'(958)$ becomes predominantly a $gg$ glue 
state. This is problematic and not in line with PDG, which 
interprets $\eta'(958)$ as predominantly a $q\bq$ state. 

Moreover, experimentally we have \cite{pdg}
\bea
&J/\Psi \rightarrow \gamma \eta'(958)
\simeq (5.15 \pm 0.16) \times 10^{-3},  \nn\\
&J/\Psi \rightarrow \gamma \eta(548) 
\simeq (1.104 \pm 0.034)\times 10^{-3},  \nn
\eea
so that we expect
\bea
&R\big(\eta(548)/\eta'(958)\big)\simeq 0.21 \pm 0.01.
\label{rratio}
\eea
But Table VI implies that this is very small
\bea
&R\big(\eta(548)/\eta'(958)\big)\simeq 0.01.
\eea
This does not agree with PDG. This again is because 
the numerical analysis interprets $\eta'(958)$ to be 
predominantly a glue state.

We could choose different input. But with $\eta(548)$,
$\eta'(958)$, and $\eta(1760)$ as the input we obtain 
very similar result. In this case the fourth physical state 
becomes $\eta(1405)$, and $\eta'(958)$ remains 
predominantly a two chromon bound state. So we have 
the same problem.
   
In this section we have discussed the numerical analysis 
of the quark gluon mixing in three channels $0^{++}$, 
$2^{++}$, ans $0^{-+}$ below 2 GeV based on our quark 
and chromon model. Clearly the numerical result is 
inconclusive and should be viewed as preliminary. 

In the $0^{++}$ and $2^{++}$ channels the numerical 
results seems to work, but in the $0^{-+}$ channel it
has problem. On the other hand we emphasize that the 
above numerical analysis is not intended to provide a 
perfect mixing. Obviously it is a rough approximation 
which is expected to have uncertainty of at least 20 
to 30 \%.

There are many reasons the above analysis can not be
perfect. First of all, the mixing discussed here neglected 
many things. For example, we have neglected the light hybrid 
$q\bq g$ states which could influence the mixing very much. 
Moreover, the mixing depends on the input parameters, but 
there are many ways to choose the input. So we have to have 
a more thorough numerical analysis.

Nevertheless, our mixing analysis confirms the followings. 
First, our quark and chromon model provides a conceptually 
simple way to identify the glueballs. Second, the mixing 
influences the physical contents of hadrons very much. 
This makes the mixing analysis more important. 

An important outcome of the analysis is that the constituent 
mass of the chromon is around several hundred MeV. This 
seems to agree with the lattice result \cite{lqcd1,lqcd2}. 

\section{Hybrids}

The quark and chromon model predicts the hybrid hadrons 
made of quark and chromon. Clearly we can construct color 
singlet $q\bq g$ mesons with one color octet chromon and 
a $q\bq$ octet. Similarly we can have $qqq g$ baryons 
with one chromon and a $qqq$ octet. So these hybrids must 
exist.

Of course, similar hybrid hadrons or multi-quark hadrons
have also been proposed before \cite{ish,chan}. But our 
model provides a unique picture of hybrid hadrons which 
is different from the other models of hybrids or multi-quark 
hadrons. In particular, it has unambiguous predictions and 
can in principle easily be distinguished from the other 
existing hybrids and/or multi-quark models. 

To understand this, notice that on the surface our $q\bq g$ 
hybrid mesons might look very similar to $qq\bq \bq$ states, 
because the chromon in $q\bq g$ could be replaced by a 
$q\bq$ octet. However, there is a clear difference between
the tetra-quark states and our $q\bq g$ hybrids. Obviously 
the $q\bq$ forms octet and singlet, but our chromon has 
no singlet component. So the spectrum (i.e., the number 
of states) that they predict is different. In other words
our hybrid model predicts less physical states. 

Similarly our $qqqg$ hybrid baryons could be misidentified 
as $qqqq\bq$ penta-quark states. But again the group 
theoretic structure of the two models is different. This  
confirms that the hybrids predicted by our quark and chromon 
model is different from other hybrids or multi-quark 
models. This means that by studying the spectrum we can 
tell which model is correct.

An important difference of these hybrids from the glueballs 
is that the hybrids have no intrinsic instability. This is 
because the chromon in $q\bq g$ and $qqqg$ hadrons 
is stable, since there is no way that it can annihilate. So, 
unlike the glueballs, these hybrids are expected to have 
typical hadronic decay width.

What is really remarkable about our hybrid hadrons is 
that it is based on the quark and chromon model. It is a 
straightforward generalization of the quark model which 
comes from the existence of the valence gluons, and 
the physics behind it is as simple as the quark model.
This simplicity translates to the clarity of the prediction.
The predictions are straightforward and unambiguous.  
So we can easily qualify or disqualfy the model experimentaly.   
This is a most important feature of our hybrid model. 

The remaining task is to identify the hybrid hadrons. 
Of course, PDG has already accumulated enough data 
which could be interpreted as hybrid hadrons and/or 
multi-quark hadrons. For example, there are quite many
low-lying mesons which can not be easily explained by 
the quark model, and some of them colud be interpreted 
as a $q\bq g$ hybrid.  So we have to analize these data 
carefully to find which model can correctly explain these 
data. This task will be tedious and time consuming, but 
certainly worth to do.

The XYZ particles might be interesting candidates of 
the hybrids \cite{xyz1,xyz2}. These particles has been 
interpreted as tetra-quarks mesons or meson-meson 
molecular bound states, but it would be worth to 
see if they could also be understood as the $q\bq g$
hybrids. 

As we have remarked, the hybrid hadrons can influence 
the quarkonium-glueball mixing significantly. So 
understanding these hybrids is very important in
the analysis of the mixing. 

\section{Monoball: Vacuum Fluctuation of Monopole Condensaton}

QCD generates the monopole condensation (more precisely 
the monopole-antimonopole pair condensation) which induces 
the dimensional transmutation and creates the mass gap. 
If so, one may ask what (if any) is the observable 
consequence of the monopole condensation. The answer 
could be the monoball.

To understand this, consider the ordinary superconductor 
in QED. It is well known that the BCS superconductivity is 
charactrized by two scales, the correlation length of the 
Cooper pair and the penetration length of the magnetic 
field. Field theoretically they are represented by two 
composite fields, a (complex) scalar field for the Cooper 
pair and a (massive) vector field for the confined magnetic 
field. And the existence of these modes are the consequence 
of the BCS superconductivity. 

So in QCD we may expect a similar consequence of the 
monopole condensation. Naively we might think that
the monopole condensation creates two mass scales, 
the correlation length of the monopole-antimonopole 
pairs and the penetration length of the chromo-electric 
flux. This suggests that the monopole condensation 
could induce two physical states, one $0^{++}$ and one 
$1^{++}$ vacuum fluctuation modes \cite{prl81,prd81}.  

However, the confinement in QCD is not exactly dual 
to the superconductivity in QED. First of all, in 
QED the magnetic field to be confined is generated 
by the electric current, not by the magnetic charge. 
But in QCD the colored flux to be confined comes from 
the color charge, not the chromo-magnetic current. 
Second, in superconductor the Cooper pair has electric 
charge but the monopole-antimonopole pair in QCD obviously 
has no chromo-magnetic charge. 

Third, in the superconductor the magnetic field is actually
screened by the supercurrent, not confined by the Cooper 
pair. But in QCD the chromo-electric field is confined by the 
monopole-antimonopole pair. In other words, it is not the 
monopole supercurrent which provides the confinement. 
QCD has no monopole supercurrent. Fourth, the chormo-electric 
flux is described by the Coulomb (i.e., scalar) potential, not 
by the vector potential, in QCD. But in superconductor the 
magnetic field is described by the vector potential. 

Finally, in the superconductor the Higgs mechanism takes 
place. The Landau-Ginzburg theory of superconducivity is 
a classic example of Higgs mechanism, where the spontaneous 
symmetry breaking generates the massive vector field which 
screens the magnetic field. But in QCD there is no spontaneous 
symmetry breaking. It is the dynamical symmetry breaking 
which generates the confinement. This tells that the confinement 
mechanism in QCD is not exactly dual to the Meissner effect. 
They are different.

In particular, this implies that the penetration length in QCD
could be represented by a scalar field, not by a spin-one field. 
This is because the chromo-electric field is described by 
the Coulomb (i.e., scalar) potential. This strongly suggests 
that both the correlation length and the penetration length 
in QCD must be represented by the scalar mode. In other 
words, there might be no $1^{++}$ vacuum fluctuation 
mode in QCD. 

The remaining question is if the two scalar modes are different 
or not. In principle they could be different, but as we have 
shown in (\ref{mgap}) the monopole condensation generates 
only one mass scale. This, together with the fact that QCD has 
only one scale $\Lambda_{QCD}$, strongly suggests that 
they are the same. 

From this we may conclude that the monopole condensation 
could have only one $0^{++}$ vacuum fluctuation mode which 
could naturally be called the magnetic glueball or simply 
the monoball. Clearly this fluctuation mode must be different 
from the glueballs made of the chromons because this 
characterizes the monopole condensation.

The importance of the monoball is that this represents the
monopole condensation, so that the experimental varification
of this monoball can be interpreted as the confirmation 
of the monopole condensation. This makes the experimental 
identification of the monoball a most urgent issue in QCD. 

Ultimately, however, the nature of the monopole condensation 
(and the number the vacuum fluctuation modes) should be 
determined by experiment, and it could well be that the 
monopole condensation has no vacuum fluctuation mode 
at all. To understand this possibility consider the Dirac 
sea, the vacuum of Dirac's theory of electron. It has vacuum 
bubbles made of electron-positron pairs, but is not the 
electron-positron pair condensation and apparently has 
no fluctuation mode. 

So, if the QCD vacuum is like the Dirac sea, there will 
be no vacuum fluctuation and thus no monoball. At the moment 
it is not clear if the QCD vacuum is similar to Dirac 
sea, and only experiments can tell whether the nature 
of the QCD vacuum is different from the Dirac sea or not. 
This makes the experimental confirmation of the monoball 
more interesting.

One might ask if there is any candidate of the monoball. 
Actually PDG has several isoscalar $0^{++}$ states, in 
particular $f_0(500)$ and $f_0(980)$, which do not fit 
well in the quark model. It would be very interesting to 
find which of them (if at all) could be interpreted as the 
monoball. 

Finally, it goes without saying that this vacuum fluctuation 
(if exists) could influence our analysis of the mixing in 
the $0^{++}$ channel. This is another complication we have 
to keep in mind in discussing the mixing. 

\section{Discussions}

In this paper we have discussed the hadron spectrum of 
the ECD obtained from the Abelian decomposition of QCD.
Although ECD is mathematically identical to QCD, it makes
the hidden dynamical structures of QCD explicit. In particular,
it assures the existence of two types of gluons and generalizes 
the quark model to the quark and chromon model. 

To compare this with other glueball models, consider the 
bag model which identifies the glueball as the colored 
field confined in a bag. In this picture the glueballs are 
made of infinite number of gluons in the form of the gluon 
field, so that there is no constituent gluon (i.e., a finite 
number of gluons) which make up the glueballs.

In contrast in the constituent model the glueballs are 
made of the constituent gluons. To bind the constituent 
gluons, however, we certainly we need the binding gluons 
(i.e., the gluon field). Unfortunately this model does 
not tell how to distinguish the binding gluons from 
the constituent gluons.      

The Abelian decomposition tells how to resolve this difficulty.
It tells that there are indeed two types of gluons which play 
different roles, and naturally generalizes the quark model to 
the quark and chromon model. This provides a new picture 
of glueballs made of chromons. Moreover, this predicts new 
hybrid hadronic states which are made of quarks and chromons. 

One of the main problems in hadron spectroscopy has been 
the identification of the glueballs. This identification 
is not simple for the following reasons. First, the glueballs 
have intrinsic instability which comes from the asymptotic 
freedom and anti-screening. Moreover, the glueballs in general 
may not exist as mass eigenstates because of the mixing with 
quarkoniums and other light hybrid mesons. 

In this paper we have discussed how to identify them by 
discussing the glueball-quarkonium mixing in the numerical 
analysis. Clearly the mixing discussed here is a rough 
approximation, because it neglects the hybrids made of
$q\bq g$ which could influence the mixing. Besides, the 
analysis depends on the input parameters, and there are 
many possibility of choosing the input which we did not 
discussed in this paper. Nevertheless it tells that the 
quark and chromon model provides a new picture of 
glueball-quarkonium mixing which can easily tested 
by experiments.       
  
In $0^{++}$ channel, our analysis is in line with (or at 
least not in contradiction with) PDG interpretation. It 
implies that $f_0(1500)$ could be predominantly the glue 
state. But here we must know which one, $a_0(980)$ or 
$a_0(1450)$, we should treat as the iso-triplet partner 
of the iso-singlet $q\bq$ which mixes with the glueball. 
This is a very sensitive question, because the numerical 
analysis depends very much on this. PDG suggests $a_0(1450)$ 
to be the iso-triplet partner. But this seems against 
the common sense, because $K_0^*(1430)$ made of $s\bar s$ 
becomes lighter than $a_0(1450)$. Clearly this issue 
remains to be settled.  

Moreover, in this channel $f_0(500)$ and $f_0(980)$ 
have been puzzling \cite{pdg}. For example, $f_0(500)$ 
has unusually broad width, and has been the subject 
of a large number of theoretical works. It has been 
suggested to be a tetra-quark state or $K \bar{K}$ 
molecules \cite{jaffe2,wein,oll,gomez}. Unfortunately 
our analysis does not reveal much about their content.

In $2^{++}$ channel our analysis could explain the physical 
content of $f_2(1270)$, $f_2'(1525)$, and $f_2(1950)$ quite 
well. In particular it could predict the relative ratio of 
the $\Psi$ radiative decay. This is remarkable. But we have
to keep in mind that there are many other so-called unconfirmed 
physical states below 2 GeV in this channel, and they have 
to be studied more carefully. 

Finally in $0^{-+}$ channel, our mixing analysis was problematic.
It implies that $\eta'(958)$ is predominantly two glue state, 
but this view is against the PDG suggestion. On the other hand, 
it is well known that this channel has a long history of problem, 
and even the origin of the octet-singlet mixing in this channel 
has not been completely understood yet. Moreover, the existence 
of a light glueball made of three chromons makes the situation 
worse. So it is natural that our mixing analysis is least 
successful. To clarify these complications we certainly need 
a more thorough analysis.

Independent of the details, however, we emphasize the 
conceptual simplicity and clarity of the quark and 
chromon model. ECD makes QCD simple by decomposing 
it to the restricted part which describes the core 
dynamics of QCD and the valence part which represents 
the colored source of QCD. This provides the clear 
picture of the glueballs and hybrid hadrons. Moreover, 
this provides a clear picture of the glueball-quarkonium 
mixing. 

In particular, ECD allows us to demonstrate the monopole 
condensation, more precisely the monopole-antimonopole 
pair condensation \cite{prd13,ijmpa14}. In this paper 
we have discussed how to verify this monopole condensation 
experimentally by searching for the monoball, the $0^{++}$ 
vacuum flucuation of the monopole condensation. 

The monoball, if exist, could have mass around $\Lambda_{QCD}$.
This implies that $f_0(500)$ could be the monoball candidate. 
Of course, at the moment it is not clear if this is the case. 
But the search for the monoball should be treated as one of 
the most important issue in QCD, because this could confirm 
the monopole condensation in QCD. 

The main purpose of this paper was to provide the general
framework of the glueball-quarkonium mixing mechanism. We 
hope to provide a more complete numerical mixing analysis 
in a separate publication \cite{imp}.

\section{Appendix}

In the Appendix we summarize some useful data for our analysis
from the Particle Data Group Review.

\begin{table*}[htbp]
\label{qmodel}
\caption{Suggested $q\bq$ quark model interpretation
of the light meson states from PDG. In the table the 
classification of the $0^{++}$ mesons is supposed to 
be tentative.} 
\begin{tabular}{|cc|c|c|c|c|c|c|}
\hline \hline $1~^{2s+1}l_J$ & $J^{PC}$ & I=1 &
I=$\frac{1}{2}$ & I=0 & I=0 & I=0 & I=0  \\
 & & $u\bar{d},~\bar{u}d,~(d\bar{d}-u\bar{u})/\sqrt{2}$ & 
 $u\bar{s},~d\bar{s},~\bar{d}s,~\bar{u}s$ & 
 $f'$ & $f$ & $c\bar{c}$ & $b\bar{b}$  \\
\hline $1~^1S_0$ & $0^{-+}$ & $\pi$ & $K$ & $\eta(548)$ & 
$\eta'(958)$ & $\eta_c(2984)$ & $\eta_b(9398)$ \\
\hline $1~^3S_1$ & $1^{--}$ & $\rho(770)$ &
$K^*(892)$ & $\phi(1020)$ & $\omega(782)$ & 
$J/\psi(3097)$ & $\Upsilon(9460)$ \\
\hline $1~^1P_1$ & $1^{+-}$ & $b_1(1235)$ &
$K_{1B}$$^{\dag}$ & $h_1(1380)$ & $h_1(1170)$ & 
$h_c(3525)$ & $h_b(9899)$ \\
\hline $1~^3P_0$ & $0^{++}$ & $a_0(1450)$ &
$K^*_0(1430)$ & $f_0(1710)$ & $f_0(1370)$ & 
$\chi_{c0}(3415)$ & $\chi_{b0}(9859)$ \\
\hline $1~^3P_1$ & $1^{++}$ & $a_1(1260)$ &
$K_{1A}$$^{\dag}$ & $f_1(1420)$ & $f_1(1285)$ & 
$\chi_{c1}(3511)$ & $\chi_{b1}(9893)$ \\
\hline $1~^3P_2$ & $2^{++}$ & $a_2(1320)$ &
$K_2^*(1430)$ & $f_2'(1525)$ & $f_2(1270)$ & 
$\chi_{c2}(3556)$ & $\chi_{b2}(9912)$ \\
\hline $1~^1D_2$ & $2^{-+}$ & $\pi_2(1670)$ &
$K_2(1770)$$^{\dag}$ & $\eta_2(1870)$ & $\eta_2(1645)$ &   &   \\
\hline $1~^3D_1$ & $1^{--}$ & $\rho(1700)$ &
$K^*(1680)$ &  & $\omega(1650)$ & $\psi(3770)$ &  \\
\hline $1~^3D_2$ & $2^{--}$ &  &
$K_2(1820)$ &   &   &   &   \\
\hline $1~^3D_3$ & $3^{--}$ & $\rho_3(1690)$ &
$K_3^*(1780)$ & $\phi_3(1850)$ & $\omega_3(1670)$ &   &   \\
\hline $2~^3S_1$ & $1^{--}$ & $\rho(1450)$ &
$K^*(1410)$ & $\phi(1680)$ & $\omega(1420)$ &   &   \\
\hline \hline
\end{tabular} 
\label{qqbarmeson}
\end{table*}

\begin{table*}[htbp]
\label{noqmeson}
\caption{Low-lying iso-singlet mesons which do not fit easily 
in the quark model listed by PDG.}
\begin{tabular}{|cc|c|c|c|c|}
\hline \hline States & $J^{PC}$ & Mass (MeV) & 
Width (MeV) & Decay modes & Branch ratio ($\%$)  \\
\hline $f_0(500)$ & $0^{++}$ & $400 \sim 550$ & 
$400 \sim 700$ & $\pi \pi$ & dominant  \\
&  &  &  & $\gamma \gamma$ & seen \\
\hline $f_0(980)$ & $0^{++}$ & $990 \pm 20$ & 
$40 \sim 100$ & $\pi \pi$ & dominant  \\
&  &  &   & $K\bar{K}, \gamma \gamma$ & seen  \\
\hline $\eta(1295)$ & $0^{-+}$ & $1294 \pm 4$ & 
$55 \pm 5$ & $ \eta \pi \pi, a_0(980) \pi$ & seen  \\
\hline $\eta(1405)$ & $0^{-+}$ & $1409 \pm 2$ & 
$51 \pm 3$ & $K \bar K \pi, \eta \pi \pi$ & seen  \\
&  &   &   & $a_0(980) \pi, 4\pi, \rho\rho$ & seen\\
\hline $\omega(1420)$ & $1^{--}$ & $1400\sim 1450$ & 
$180-250$ & $ \rho \pi$ & dominant  \\
&  &   &   & $\omega \pi \pi$ & seen  \\
\hline $\eta(1475)$ & $0^{-+}$ & $1476 \pm 4$ & 
$85 \pm 9$ & $K \bar K \pi$ & dominant  \\  
&  &   &   & $a_0(980) \pi, \gamma \gamma$ & seen\\
\hline $f_0(1500)$ & $0^{++}$ & $1505 \pm 6$ & $109 \pm 7$ & 
$\pi \pi$ & $34.9 \pm 2.3$   \\
&  &  &  & $4\pi$ & $49.5 \pm 3.3$   \\
&  &  &  & $K\bar{K}$ & $8.6 \pm 1.0$  \\ 
&  &  &  & $\eta \eta$ & $5.1 \pm 0.9$ \\
&  &  &  & $\eta \eta'(958)$ & $1.9 \pm 0.8$  \\
\hline $\eta_2(1645)$ & $2^{-+}$ & $1617 \pm 5$ & 
$181 \pm 11$ & $a_0(980) \pi, a_2(1320) \pi, K^* \bar K$ & seen  \\  
& &   &   & $K\bar K \pi, \eta \pi^+ \pi^-$ & seen\\
\hline $\phi(1680)$ & $1^{--}$ & $1680\pm 20$ & $150\pm 50$
& $K\bar K^*(892)$ & dominant  \\ 
&  &  &  & $k \bar K, e^+e^-$ & seen  \\
\hline $f_2(1950)$ & $2^{++}$ & $1944 \pm 12$ & 
$472 \pm 18$ & $K \bar K, K^*(892) \bar{K}^*(892)$ & seen  \\
& &   &   & $\pi \pi, 4\pi, \eta\eta, \gamma\gamma$ & seen  \\
\hline \hline
\end{tabular} 
\label{qqbarmeson}
\end{table*}

{\bf ACKNOWLEDGEMENT}

The work is supported in part by the National Natural 
Science Foundation of China (Grants 11175215, 11447105, 
and 11475227), Chinese Academy of Sciences Visiting
Professorship for Senior International Scientists 
(Grant 2013T2J0010), Basic Science Research Program 
through the National Research Foundation of Korea (NRF) 
funded by the Ministry of Science and Future Planning 
(Grant 2012-002-134), and by Konkuk University.

\end{document}